\documentclass[sigconf]{acmart}
\usepackage{afterpage}  

\usepackage[normalem]{ulem}
\usepackage{longtable}

\newcommand{\pains}{A}

\definecolor{burgundy}{rgb}{0.5, 0.0, 0.13}

\AtBeginDocument{%
  }

\setcopyright{acmlicensed}
\acmYear{2025}\copyrightyear{2025}
\acmConference[ACM FAccT '25]{ACM Conference on Fairness, Accountability, and Transparency}{June 23--26, 2025}{Athens, Greece}
\acmBooktitle{ACM Conference on Fairness, Accountability, and Transparency (ACM FAccT '25), June 23--26, 2025, Athens, Greece}\acmDOI{}
\setcopyright{none}
\renewcommand\footnotetextcopyrightpermission[1]{}

\begin{document}

\title{From tools to thieves: Measuring and understanding public perceptions of AI through crowdsourced metaphors}

\author{Myra Cheng}
\authornote{Both authors contributed equally to this research.}
\email{myra@cs.stanford.edu}
\affiliation{%
  \institution{Department of Computer Science, Stanford University}
  \city{Stanford}
  \state{CA}
  \country{USA}
}

  \author{Angela Y. Lee}
\authornotemark[1]
\email{angela8@stanford.edu}
\affiliation{%
  \institution{Department of Communication, Stanford University}
 \city{Stanford}
  \state{CA}
  \country{USA}}
  \author{Kristina Rapuano}
\affiliation{%
  \institution{BetterUp}
  \city{San Francisco}
  \state{CA}
  \country{USA}}
  \author{Kate Niederhoffer}
\affiliation{%
  \institution{BetterUp}
  \city{San Francisco}
  \state{CA}
  \country{USA}}
  \author{Alex Liebscher}
\affiliation{%
  \institution{BetterUp}
  \city{San Francisco}
  \state{CA}
  \country{USA}}
\author{Jeffrey Hancock}
\affiliation{%
  \institution{Department of Communication, Stanford University}
 \city{Stanford}
  \state{CA}
      \country{USA}}

\renewcommand{\shortauthors}{Cheng et al.}

\begin{abstract}
How has the public responded to the increasing prevalence of artificial intelligence (AI)-based technologies? We investigate public perceptions of AI by collecting over 12,000 responses over 12 months from a nationally representative U.S. sample. Participants provided open-ended metaphors reflecting their mental models of AI, a methodology that overcomes the limitations of traditional self-reported measures by capturing more nuance. Using a mixed-methods approach combining quantitative clustering and qualitative coding, we identify 20 dominant metaphors shaping public understanding of AI. To analyze these metaphors systematically, we present a scalable framework integrating language modeling (LM)-based techniques to measure key dimensions of public perception: anthropomorphism (attribution of human-like qualities), warmth, and competence. We find that Americans generally view AI as warm and competent, and that over the past year, perceptions of AI's human-likeness and warmth have significantly increased ($+34\%, r = 0.80, p < 0.01;  +41\%, r = 0.62, p < 0.05$). These implicit perceptions, along with the identified dominant metaphors, strongly predict trust in and willingness to adopt AI ($r^2 = 0.21, 0.18, p < 0.001$). Moreover, we uncover systematic demographic differences in metaphors and implicit perceptions, such as the higher propensity of women, older individuals, and people of color to anthropomorphize AI, which shed light on demographic disparities in trust and adoption.
In addition to our dataset and framework for tracking evolving public attitudes, we provide actionable insights on using metaphors for inclusive and responsible AI development.

\end{abstract}

\received{20 February 2007}
\received[revised]{12 March 2009}
\received[accepted]{5 June 2009}

\maketitle

\section{Introduction}

Advances in large language models (LLMs) have catalyzed public interest in artificial intelligence (AI) \cite{grossmann2023ai,demszky2023using,jan2023artificial,chavali2024ai,financialtimes2024,hine2023blueprint,boine2023emotional}, 
with over 90\% of Americans having heard of AI \cite{pewresearchcenter_2023_viewofai} and tools like ChatGPT now receiving over one billion queries per day \cite{roth2024chatgpt}.
Narratives of AI have been shaped by both hope and fear as well as trust and skepticism of its societal impacts \cite{sartori2023minding, glikson2020human}. 
While some view AI as a transformative tool for productivity \cite{hazarika2020artificial,eubanks2022artificial,stade2024large} or as a trusted assistant for personal and professional tasks \cite{danaher2018assistant}, others see it as a threat to their livelihoods \cite{kochhar2023us}. Perceptions of AI are important determinants of individual behavior, industry-level changes, and community support for regulatory measures \cite{gilardi2024we}. For example, an individual's decision to use AI to answer their questions, process their emotions, or assist them with their work is highly dependent on their perception of the system's capabilities \cite{kelly2023factors}. Decisions to regulate AI are also fundamentally shaped by perceptions about these systems \cite{wilczek2024government}. Understanding public perceptions of AI is therefore integral to grasping AI's societal impacts.

Perceptions about new technologies, however, can be difficult to capture using traditional survey measures \cite{gallagher2012health,brugman2017recategorizing, jiang2021supporting, jensen2024reflection}. People often struggle to verbalize their nuanced perceptions of complex sociotechnical systems like AI, particularly when they have less experience with them \cite{lee2021role}. To overcome these limitations, we study public perceptions of AI by analyzing the \textit{metaphors} they use to describe AI \cite{demir2022determining}. People have long used metaphors to communicate complex ideas, like the turbulence of emotions (``rollercoaster of feelings'') or the value of digital resources (``data is the new oil'') \cite{lakoff1980metaphors}. We conducted an online study where participants were asked to provide their metaphors of AI, collecting responses from a nationally representative sample of over 12,000 Americans from May 2023 (shortly after the mainstream adoption of ChatGPT \cite{wu2023brief}) to August 2024. The collection of these metaphor data over this critical period provides insights into how the public interprets and understands the rise of generative AI and public-facing tools.

First, we identified the \textit{dominant metaphors} people used throughout this time period to describe AI (e.g., ``AI is a tool'', ``AI is a thief'') using a combination of automatic clustering and qualitative refinement.
Beyond the explicit descriptions of the metaphors, we also demonstrate how these metaphors provide a window into people's \textit{implicit perceptions of AI}. We measure implicit perceptions from the metaphors in two ways: 1) the extent to which participants anthropomorphized AI, using a method measuring implicit framings from language model (LM)-based probabilities \cite{cheng-etal-2024-anthroscore}, and 2) how warm and competent they perceive AI, using semantic axes constructed with LM-based embeddings \cite{fraser-etal-2021-understanding}. These constructs are important predictors of people's experiences with AI: Anthropomorphism, or the attribution of human-like qualities to AI, has come to the fore of discussions around societal impacts of AI \cite{salles2020anthropomorphism,shanahan2023role,troshani2021we,cheng2025iamthe} as scholars have highlighted the consequences of users becoming overly dependent on human-like AI \cite{abercrombie-etal-2023-mirages,chiesurin-etal-2023-dangers,chan2023harms,Inie2024-dy} and failing to consider the implications of disclosing sensitive information \cite{Ischen2020-it,citron2022privacy}. Warmth and competence are integral dimensions of how people form impressions of social actors \cite{fiske2007universal, mieczkowski2019helping}.
The extent to which people believe that others have warm, positive intentions and are capable of carrying out given tasks are strong predictors of people's attitudes, such as their trust in new technologies and willingness to adopt them in the future \cite{mckee2023humans,li2024warmth,khadpe2020conceptual,zhou2024rel}.

Using this framework to analyze our dataset, we find that while participants' metaphors are generally warm and competent, they vary widely in anthropomorphism (e.g., ``AI is a friend'' versus ``AI is a library''). We also identify significant shifts over time in implicit perceptions: anthropomorphism and warmth have significantly increased by $34\%^{**}$ and $41\%^*$\footnote{Throughout the paper, we use *, **, ***, and (ns) to denote $p < 0.05, p < 0.01,$ $p < 0.001$, and no statistical significance respectively.} respectively, over this time period.

We next demonstrate how dominant metaphors and implicit perceptions predict attitudes toward AI: warmth, competence, and anthropomorphism, as well as the dominant metaphors of ``god'', ``brain,'' and ``thief'', are the most predictive of trust and adoption (``thief'' being negatively predictive).  

We also discover significant demographic differences in the dominant metaphors, implicit perceptions, and attitudes Americans have about AI. For instance, gender differences in metaphors about AI help provide insight into why women may trust it less than men \cite{pewresearchcenter_2023_viewofai}. Surprisingly, we also find that Americans of color and older Americans tend to trust AI more -- indicating the importance of investigating how social identities shape human-AI interactions. These differences inform efforts to preemptively recognize and address adverse effects of AI on marginalized communities.

\paragraph{Summary of Contributions.} Our study provides a fine-grained understanding of how public perceptions of AI have emerged over time and across individuals. We collect a dataset of over 12,000 metaphors of AI over 12 months and use mixed methods to identify 20 dominant metaphors. We provide a scalable framework for measuring implicit perceptions from the metaphors. We use this framework to demonstrate how temporal shifts in metaphors and implicit perceptions correspond to changes in people's attitudes, and explore how demographic differences among these reveal harmful consequences of AI deployment. Our framework\footnote{Our code is available at \url{https://github.com/myracheng/ai_metaphors}, and our dataset is available at \url{https://osf.io/jers8}.} is quantitative and scalable, and thus can be easily adapted to new data to capture evolutions in public perception \cite{fast2017long,zhang2019artificial}. 

\section{Background}

\subsection{Using metaphors to reveal implicit perceptions}
The metaphors people use to describe AI can reveal implicit perceptions that may be difficult for them to verbalize explicitly \cite{gallagher2012health,brugman2017recategorizing}. People often struggle to accurately articulate their specific attitudes and perceptions when asked to provide their responses to new technologies.
For instance, individuals’ self-reported perceptions of anthropomorphic qualities often fail to align with their implicit beliefs or behaviors—such as treating computers as social actors but not realizing that they perceive computers as human-like \cite{zlotowski2018model}. 
Thus, to access these underlying perceptions, researchers use methodologies such as asking participants to answer indirect questions or provide metaphors \cite{demir2022determining}. 
Similarly, we use metaphors to conduct a large-scale study of people’s perceptions of AI.
\subsection{Why metaphors matter}
Metaphors shape how people understand information by distilling complex concepts into more accessible ideas \cite{brandt2005making, lakoff1980metaphors}. People use metaphors to make abstract ideas feel tangible by describing them using the language and properties of more familiar concepts \cite{zeilig2022dementia}. The implicit associations invoked by these comparisons can have powerful effects by orienting people towards specific conceptualizations and behaviors. Decades of psychological research demonstrates that the metaphors people use to understand abstract concepts, like \textit{crime}, \textit{illness}, and \textit{intelligence}, can unconsciously change their behaviors \cite{hendricks2018emotional, thibodeau2013natural}. For example, those who viewed local crime as a ``virus'' infecting their city were more likely to empathize with its perpetrators than those who viewed crime as a ``beast'' preying on their city; framing crime as a virus caused people to endorse rehabilitative policies over punitive measures \cite{tversky1974judgment,thibodeau2016extended}.

Because they help people make sense of unfamiliar concepts, metaphors have long been used to help people understand new technologies. Early metaphors of the Internet as a ``superhighway'' that could connect users to diverse digital destinations \cite{fluckiger1996world} and the later metaphor of ``surfing the web'' that suggests the Internet as a vehicle for exploration \cite{maglio1998metaphors}.
Similarly, understanding the metaphors that people use to describe AI may reveal how the public is responding to the proliferation of AI-based technologies, and how they understand the potential values and uses that AI technologies may afford. Clear differences are emerging in how people think about AI. For instance, some people describe AI is a powerful tool that should be integrated into the workforce \cite{hazarika2020artificial,eubanks2022artificial,stade2024large}. Others view AI as an assistant or companion that can be trusted to process personal and professional challenges \cite{danaher2018assistant}. However, many people view AI to be a threat to their livelihoods, invoking fears of replacement and may therefore instead advocate for restrictions on AI \cite{kochhar2023us}. Previous work has also demonstrated 
 the impact of human versus non-human metaphors in conversational agent design \cite{desai2023metaphors, jung2022great}. The metaphors people use to describe AI provide insight into how people feel about using it in their own lives, in the context of their work, and in society at large. 

\subsection{Implicit perceptions of AI}\label{sec:percofai}
People's implicit perceptions play a powerful role in shaping human interactions with technology. Even when people interact with the same system, they can interpret the qualities of the system in vastly different ways that affect their trust and engagement. Understanding these perceptual differences is vital for  building and deploying AI responsibly. 
Because the inner workings of these systems are often ``black boxes'', individuals rely on their intuitive beliefs when communicating with and through AI \cite{ytre2021folk}. Similarly, the effects of interacting with AI depend on how people perceive and interpret their experiences with these complex socio-technical systems. Previous research on generative AI has emphasized the importance of considering key dimensions of perceptions, which we measure using our quantitative framework:

\textit{1) Anthropomorphism:}
Anthropomorphism, or the attribution of human-like characteristics in a non-human entity, spans a broad range of qualities, including cognitive, behavioral, and emotional traits \cite{epley2007seeing, devrio2025taxonomy,cheng2025humt,cheng2025dehumanizing}. In the context of AI, this can manifest in applications like conversational assistants \cite{gabriel2024ethics}, relationship-building chatbots \cite{brandtzaeg2022my}, and AI-generated simulations of individuals \cite{mcilroy2022mimetic,olteanu2025ai}. 
Anthropomorphism of AI is contentious at the individual and societal levels, as it raises concerns around assigning moral responsibility to AI \cite{friedman2007human}, overreliance on AI \cite{Watson2019-py,Zarouali2021-gy}, and degrading emotional and social relations \cite{akbulut2024all,Maeda2024-cv,Shteynberg2024-cg,laestadius2022too}.

\textit{2) Warmth and Competence:}
One of the fundamental ways that people perceive others is in terms of their warmth and competence. Warmth refers to the extent that an entity is seen as  friendly, trustworthy, kind, and caring. Competence refers to an entity's ability to act and captures beliefs about whether they are capable and intelligent. These dimensions are not only applied to humans \cite{cuddy2009stereotype} but also extend to non-human entities, including robots and other technologies \cite{mieczkowski2019helping}, and impact attitudes towards technology, e.g., people prefer AI agents that seem less competent \cite{khadpe2020conceptual}, and people rely more on AI that they perceive as warm \cite{zhou2024rel}.  
\subsection{Attitudes toward AI: Trust and adoption}
We focus on two core aspects of public attitudes toward AI: trust in AI and willingness to adopt AI. A certain level of trust is necessary for effective and beneficial integration of AI into society \cite{afroogh2024trust}; lack of trust and unwillingness to adopt AI may lead to under-use and missed opportunities to improve people's lives \cite{choudhury2023investigating}. However, excessive trust or premature adoption can also be dangerous, leading to over-reliance, inaccurate expectations of AI capabilities, and other harms \cite{chiesurin-etal-2023-dangers,abercrombie-etal-2023-mirages}. Over-trust can also obscure the real, ongoing harms that AI systems already impose, including the amplification of social inequalities \cite{birhane2021algorithmic, bianchi2023easily}, surveillance and erosion of privacy \cite{kalluri2023surveillance}, and the marginalization of vulnerable groups \cite{weidinger2022taxonomy, Chien2024-vl, tiku2022, hunter2023}. Moreover, public fear or backlash fueled by misplaced trust can divert attention away from addressing these concrete harms, instead focusing on speculative or exaggerated risks \cite{lukyanenko2022trust}. Understanding these variables is thus critical to shaping societal outcomes of AI.

\section{Dataset Collection Methods}

\subsection{Participants and Recruitment}
We recruited 12,933 participants from the crowdsourcing platform Prolific between May 2023 and August 2024, with approximately 1,000 individuals recruited each month, as part of a larger project understanding Americans’ experiences with AI. We note that over the 16-month period, we were unable to collect data during July 2023 and May - July 2024 due to technical issues; thus, we have responses over 12 months total. All survey procedures were approved by the BetterUp Institutional Review Board. We assessed the degree to which the samples each month were nationally representative of the United States population by using the American National Election Studies’ raking algorithms \cite{baker2013summary,pasek2016will}. The analysis revealed that our data were nationally representative of the US population with respect to gender, ethnicity, education, and age overall, as none of the variables differed by a margin of more than .5\% in any month. For full demographic details, please see Appendix \ref{sec:demo}.

\subsection{Procedure and Measures}
We elicited metaphors from participants by asking them: ``Some people use metaphors to describe abstract concepts, like AI. What is the best metaphor for how AI works?'' They were encouraged to use their own words, and were assured that there were no wrong answers. We specified that the task was not about assessing their comprehension of the details of how AI works, but rather about understanding their thoughts (``This is not about accuracy as much as understanding how you are thinking''). We chose to give this prompt with relatively little context, rather than a specific hypothetical scenario, to capture a wide range of responses. Participants entered their answer into an open text box in the form ``AI is like \_\_\_\_\_\_\_\_\_\_because\_\_\_\_\_\_\_\_\_\_\_.'' Next, participants completed a series of survey measures related to their experiences with and attitudes toward AI. We assessed the \textit{frequency of each individual's AI use} by asking them to indicate if they had heard of or used 8 of the most commonly used, consumer-facing tools: ChatGPT, DALL-E, Claude, Bard, Anthropic, Midjourney, Gemini, and Perplexity. They could also write-in the names of additional AI tools that they used. \textit{Trust in AI} was measured using a 3-item survey measure adapted from past work investigating trust in human vs. AI-generated content \cite{jakesch2019ai}, building on the Organizational Model of Trust widely used to study trust in technologies and products \cite{mayer1995integrative}. Finally, we assessed participants \textit{willingness to adopt AI} with a 5-item survey measure of their behavioral intentions to engage with AI in the future. Full details of these survey items are in Appendix \ref{sec:surv}.

\paragraph{Ensuring data quality}\label{sec:removegpt}
We took precautions to account for the possibility that people may use AI to generate inauthentic responses to our free response questions by disabling copy/paste and including attention-check questions
\cite{goodrich2023battling}. However, it may still be possible that participants used AI to generate their responses to our free-response question. To filter such responses out, we evaluated the semantic similarity of participant responses with those from ChatGPT. Full details are in Appendix \ref{sec:exclu}. We exclude 276 such participant responses, resulting in 11,790 valid metaphors.

\section{Analytic Approach}
\begin{figure*}
    \centering
    \includegraphics[width=0.97\linewidth]{ 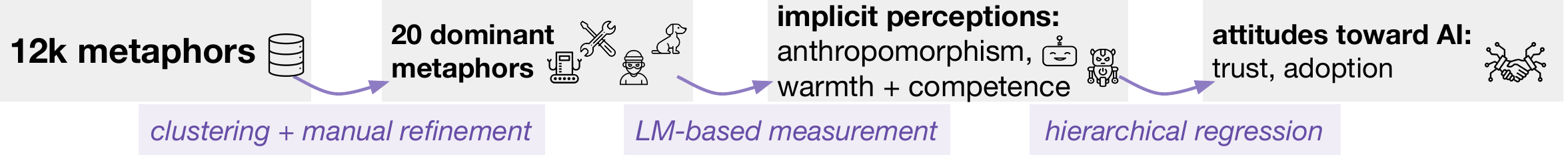}
    \caption{\textbf{Approach to analyzing our dataset of metaphors}. Our methods to analyze the metaphors include: identifying dominant metaphors from the crowdsourced dataset of metaphors; using LM-based methods to measure implicit perceptions of anthropomorphism, warmth, and competence from the metaphors; and using these to explain attitudinal variables of trust in and adoption of AI.}
    \label{fig:methods}
\end{figure*}

Our methods for large-scale analyses of the metaphors are summarized in Figure \ref{fig:methods}.

\subsection{Topic modeling to identify dominant metaphors}\label{sec:topicmodel} 
To identify thematic clusters in participants' metaphors, we use a mixed-methods approach: 1) automatic clustering using embeddings, 2) manual refinement of clusters, and 3) outlier analysis.

\textit{1) Automatic clustering:} First, we removed frequently occurring words (e.g., ``AI'' and ``artificial intelligence'') from all metaphors to ensure that the identified clusters were based on meaning rather than differences in phrasing. We then use the sentence embedding model \texttt{all-mpnet-base-v2} \cite{reimers-gurevych-2019-sentence} to generate a 768-dimensional embedding $e({m_p})$ that captures the meaning of each participant $p$’s metaphor of AI $m_p$. Following prior work \cite{kirk2024prism} and based on qualitative examination of the clusters generated from varying dimensionalities, these representations were reduced to 20 dimensions using UMAP \cite{McInnes2018} and normalized using the L2 norm to reduce complexity prior to clustering. We then clustered these representations with HDBScan, a clustering algorithm that groups similar metaphors without forcing every metaphor into a group or pre-defining a number of groups \cite{mcinnes2017hdbscan}. The clustering was performed using the BERTopic Python package \cite{grootendorst2022bertopic}. This process categorized about 80\% of the metaphors in one of 50 clusters, while the remaining 20\% were considered outliers.

\textit{2) Manual refinement:} Using iterative discussions and consensus-building methods that are standard in qualitative coding approaches to developing grounded theory \cite{oktay2012grounded,charmaz2015grounded,corbin2014basics}, the research team iteratively grouped the clusters based on thematic and conceptual similarities (e.g., combining a cluster about ``stringing together ideas''  and a cluster about ``a chef combining ingredients'' into the dominant metaphor of ``synthesizer''), focusing on distinctiveness between and coherence within clusters, until we reached consensus. This resulted in a final set of 20 clusters (Note that we did not pre-constrain the number of clusters in the final set.) We refer to these 20 most frequently occurring metaphor clusters as \textbf{dominant metaphors}. Names and descriptions for each dominant metaphor were also decided upon collaboratively.

\textit{3) Outlier analysis:} Finally, we re-assigned outlier metaphors that were not initially automatically assigned to a cluster as follows: For each dominant metaphor, we computed the centroid as the average of all the embeddings of the metaphors belonging to that cluster. Then, for each outlier metaphor $m_p$, we measured the cosine similarity of its embedding $e(m_p)$ to each centroid. $m_p$ was then categorized under the dominant metaphor with the highest cosine similarity if the similarity was greater than $0.6$.\footnote{We choose this threshold based on manual inspection of how \textit{conceptually} similar the metaphors are at different thresholds.} This process enabled us to automatically categorize metaphors that aligned with the broader themes identified in the manual refinement phase, but may have been too semantically different to be initially identified. This process results in 10,629 of the 11,789 metaphors being assigned to a dominant metaphor.

\subsection{Measuring implicit perceptions from metaphors} 
Because metaphors are laden with implicit associations \cite{gibbs2011evaluating}, they can reveal how people think about the fundamental attributes of AI, such as how human-like an AI is, or whether people view AI as part of their ``in-group.'' 

To measure implicit perceptions, we developed a scalable, systematic framework to score metaphors on the dimensions of anthropomorphism, warmth, and competence. Our approach is based on methods to automatically score any open-ended text on these dimensions, and thus we apply them to every individual metaphor. For each dominant metaphor, we compute the mean anthropomorphism, warmth, and competence score across the individual metaphors in that cluster.

\subsubsection{Measuring anthropomorphism} To assess the extent to which people’s metaphors of AI ascribed human-like characteristics to AI, we adapted AnthroScore \cite{cheng-etal-2024-anthroscore}, an automatic metric of implicit anthropomorphism in language. In their work, they use the masked language model RoBERTa to calculate the relative probability that a given entity (e.g., ``AI'') in a text would be replaced by human pronouns versus non-human pronouns. Specifically, the degree of anthropomorphism for entity $x$ in sentence $s$ is measured as

\begin{equation*}
\pains(s_x) = \log \frac{P_{\textsc{human}}(s_x)}{P_{\textsc{non-human}}(s_x)}
\text{, where} 
\end{equation*}
\begin{equation}
P_{\textsc{human}}(s_x) = \sum_{w \in \text{he, she}} P(w), P_{\textsc{non-human}}(s_x) = \sum_{w \in \text{it}} P(w),
\end{equation} where $P(w)$ is the model's outputted probability of replacing the mask with the word $w$. We apply AnthroScore to the entities $x \in \{\text{it, AI, artificial intelligence}\}$ to measure the anthropomorphism of AI in each metaphor $m_p$. If the metaphor does not contain any of these entities, we use the \texttt{spacy} package \cite{spacy2} to identify whether the metaphor is only a noun phrase. If so, we prepend the phrase ``AI is'' and then apply AnthroScore to the now-present term ``AI'' to measure anthropomorphism. 
Since this score is a relative log probability, a score greater than 0 suggests that the entity is more likely to be human, and a score less than 0 suggests that the entity is more likely to be non-human. For example, in the metaphor ``AI is a teacher'' (and the entity ``AI''), we compute the probability of the sentence ``\textit{He} is a teacher'' and the sentence ``\textit{It} is a teacher.'' Because the former is a much more probable sentence, this metaphor has AnthroScore > 0, indicating that it anthropomorphizes AI. For interpretability, we use an indicator score of whether $A(s_x)$ is greater than 0:
\begin{equation}
A^{\text{bin}}(m_p) = \mathbb{I}\big(A(m_p) > 0\big).
\end{equation}
This binary form allows us to easily and directly measure the percentage of metaphors that are anthropomorphic (i.e., $A^{\text{bin}}(m_p) = 1$), as reported in the following sections.
\subsubsection{Measuring warmth and competence.} We were also interested in understanding the extent to which people’s metaphors characterized AI as being warm and competent, two fundamental dimensions of social perception that are well-established psychological constructs for estimating perceptions of people \cite{cuddy2009stereotype} and anthropomorphized non-human entities \cite{mieczkowski2019helping}. To assess these qualities in the metaphors of AI, we follow \citet{fraser-etal-2021-understanding}'s method for constructing semantic axes for warmth and competence. This approach quantifies a text on these dimensions, by measuring the semantic similarity between a given text (in our case, a metaphor about AI) and a constructed  semantic axis that represents that dimension. We employed validated lexicons \cite{nicolas2021comprehensive} for warmth and competence as anchors to define two axes: warmth-coldness and competence-incompetence. These lexicons comprise 6180 words that have been previously validated to be strongly associated with high or low warmth and competence. Then, for each metaphor, we project its embedding onto the axis, obtaining a score between -1 and 1. This method enables us to position texts on a continuous spectrum between two extremes (e.g., warm vs. cold or competent vs. incompetent). 
Specifically, the semantic axes are defined as follows: for a given dimension $D$ (warmth or competence), we construct the semantic axis by
taking the mean of the embeddings for each word in the lexicon that are positively associated with that dimension, and subtract the mean of the embeddings for each word in the lexicon that are negatively associated with it:
\begin{equation}
    A_\text{D} = \frac{1}{k}\sum_{i=1}^k e(w_i) - \frac{1}{m}\sum_{j=1}^m e(w'_j), 
\end{equation}
where $w_i/w'_j$ is a word in the set of words positively/negatively associated with that dimension respectively. We again use\\ \texttt{all-mpnet-base-v2} to compute the embeddings.
For each metaphor $m_p$ embedded as $e(m_p)$, we compute its warmth and competence as the cosine similarity of the embedding $e(m_p)$ to this axis, i.e., \begin{equation}
\begin{split}
\text{Warmth}(m_p) &= \cos(e(m_p), A_\text{warmth}), \\
\text{Competence}(m_p) &= \cos(e(m_p), A_\text{competence})
\end{split}
\label{eq:wc}
\end{equation}resulting in two scores (each between -1 and 1) that represent the metaphor's warmth and competence respectively. Examples of metaphors with different scores for warmth and competence are in Table \ref{tab:warmthcompex}.
\begin{table*}[]\small
\begin{tabular}{|p{0.47\linewidth}|p{0.47\linewidth}|}\hline
\textbf{Positive warmth, negative competence:}

``It's like a dog. On one hand, it can be gentle and loving. However, you don't know if it may suddenly bite.''

``Like an imaginary friend''

``AI is like Your uncle is who right 50\% of time. because They have a lot to say but don't understand the context.''

``AI is like having a forever friend because they cant get up and leave''
 & \textbf{Positive warmth, positive competence:}
 
``AI is like having a ‘phone a friend’ lifeline on the TV gameshow, ‘Who wants to be a millionaire.’ You can ask for assistance from a more knowledgeble entity for personal assistance or gain.''

``AI is like a wise old friend because it knows a lot and can give very helpful advice.''

``It makes me think of a fairy god mother. All knowing and there to help you out''

``It’s like a robot that can learn over time''
\\\hline
\textbf{Negative warmth, negative competence:}

``It grinds my gears''

``Like a turtle trying to run like a rabbit.''

``Its a computer predicting the next word   It has no understanding''

"Shoddy thief''
 &\textbf{ Negative warmth, positive competence:}
 
``A detective that collects clues from different recourses, analyzes them and uses that information to make a prediction or solve a problem''

``Its a computer system that is loaded with a bunch of data. ALl the data is tied to key words so when the key word us used a bunch of data with that tag is pulled up.''

``It's like a copy machine that takes info and tries to replicate it''

``Putting human knowledge into a system and then having it spit out I got stood at you. It could be in the form or writing or even art.''\\\hline
\end{tabular}
\caption{\textbf{Metaphors with varying levels of warmth and competence (participants' responses, unedited).} We find that the vast majority of metaphors have positive warmth and positive competence.
}
\end{table*}\label{tab:warmthcompex}
\subsection{Predicting attitudes about AI from dominant metaphors and implicit perceptions}
Next, we measured how dominant metaphors and implicit perceptions help explain two attitudinal variables with consequential downstream outcomes: trust in AI and willingness to adopt AI. 
Specifically, following a model comparison approach \cite{crum2013rethinking,lee2024social}, we conducted a pair of hierarchical multiple regression analyses to examine the extent to which dominant metaphors and perceptions can explain variance in people's trust in AI and their willingness to adopt AI, relative to demographic differences and their use of AI. All predictors were standardized to [0,1] to ensure comparability of effect sizes. We first conducted a regression to explain the dependent variable (trust or adoption) using a foundational set of predictors: time, frequency of AI tool use, gender, race/ethnicity, and age. For gender, we binarize the variable into men versus non-men (women and non-binary people), and for race/ethnicity, we binarize the variable into white versus non-white. Next, we introduce the second block of variables, which consists of 20 variables each representing the presence of a dominant metaphor (tool, robot, assistant, etc.). Finally, we incorporated the third block of variables, which are the implicit perceptions measured by our framework: anthropomorphism, warmth, and competence. This sequential approach allowed us to assess the incremental explanatory power of metaphors and perceptions in predicting trust and adoption beyond demographic and usage factors. Model assumptions, including linearity, multicollinearity, and homoscedasticity, were also checked to ensure the validity of the results.

\section{Results}
\subsection{Teacher, tool, pet, or friend: dominant metaphors of AI}

As detailed in \S\ref{sec:topicmodel}, we used a combination of automatic clustering and manual iterative coding to identify the 20 dominant clusters of metaphors that people used to conceptualize AI in our dataset (Table \ref{tab:metaphors}). We find that people most commonly described AI as being like 1) a technological or analog tool, such as a calculator or Swiss Army knife (10\%), 2) a brain capable of reasoning and logic (10\%), and 3) a powerful search engine capable of navigating large databases (9\%). While these were some of the most common, our participants used a wide range of metaphors to characterize their mental models of AI. For example, some viewed AI as an intelligent teacher (``AI is like a professor because it always has the answer'', 3\%) whereas others viewed it as akin to a child (``It needs to learn, but is happy to provide output and is proud of it'', 4\%), with still others seeing AI as being like a thief (``It’s basically plagiarism'', 0.5\%). People also compared AI to various mythical beings, from folklore characters (``a fairy godmother - there to help you with whatever you need'', 1\%) to an all-knowing god (``a deity or divine creature that…can make decisions on a grand scale'', 1\%). We also saw inspiration from popular conceptions of AI from science fiction and movies, such as references to the Terminator and other robots that persist in the American imagination (``a robot that tries to think '', 8\%). Some people viewed AI as being like an inherent mechanical entity (``a software with enhanced data processing capabilities'', 7\%, or as a ``distorted mirror'', 4\%) while others  likened it to animals and pets (``It’s like training your dog to do something'', 2\%), or an unexplored realm (``AI is like the ocean, there is so much uncertainty and so much to discover'', 2\%). The wide range of dominant metaphors reveals the complexity with which people view AI. Note that small percentages potentially reflect large populations, e.g., ``thief'' appears in 0.5\% of the dataset ($n=57$) but represents millions when extrapolated to the U.S. population, where AI awareness exceeds 90\% \cite{pewresearchcenter_2023_viewofai}. 


\begin{table*}[]\tiny
\begin{tabular}{|p{0.05\linewidth}p{0.19\linewidth}p{0.56\linewidth}
p{0.01\linewidth}p{0.01\linewidth}p{0.03\linewidth}|}\hline
\textbf{Metaphor}        & \textbf{Description}                                                                                                                                 & \textbf{Examples}                                                                                          & \textbf{\%} & $n$ & $r$ \\\hline
Tool                  & A technological tool like an engine, calculator, or appliance; or an analog tool, like a hammer or Swiss Army knife & ``AI is like a very smart scientific calculator with access to the internet. It still needs input to have an accurate output. Just like how the higher end calculators work, the output may be in the wrong format and need to be adjusted in order for the data to be used.''     ``AI works like the engine on a car, the more you put into it the more you get out of it''                                                                                                        & 10\%     & 1233      & $.91$ (ns)                                         \\\hline  Brain      & A powerful brain capable of human-level thinking, intelligence, and reasoning  & ``AI is like an external brain you can access to help you solve problems''     ``AI is like a human brain but without all the information that is unnecessary or distracting. It also operates without emotions.''                                                                                                                                                                                                                                                                   & 10\%        & 1187   & $-.48$ (ns)                                                       \\\hline Search\newline engine         & A search engine or database navigator that can sift through data and information on the Internet & ``AI works similarly to Google in the sense that you can type in a question, and find your answer with the results. However, with AI, they have an entire interface that allows the AI to find what it thinks is the best answer, instead of having to research through several options.''     ``It’s like if google pretended to be a human and just spit back the top answers instead of linking you to top pages'' & 9\%      & 1029      & $-.61^*$                \\\hline Assistant        & An assistant or employee who can help users complete tasks and find answers                                                   & ``To me, AI is like having a personal assistant.  It's something that you can delegate certain tasks to be performed.  But it's not always perfect and can be prone to making mistakes.''     ``AI is like a good to average personal assistant because you can rely on it to get tasks done right... most of the time.''                                                                                                                                                            & 8\%  & 941          & $.63^*$                                                             \\\hline Robot        & A robot, sometimes embodied in a way that resembles humans                                                                       & ``AI is like robot that tries to think because it processes information but has no feelings.''     ``AI as an embodied robot that resembles humans, as in science fiction, like the Terminator.''                                                                                                                                                                                                                                                                                    & 8\% & 886           & $.72^{**}$                                    \\\hline Computer  & A powerful computer or software with enhanced data processing and analytic abilities                                  & ``AI is a computer actively thinking. It’s a computer deciding exactly how to research and respond to a prompt''
`` AI is a computerized evolution of the human mind. The ultimate goal is to create the perfect thinking and processing machine. Or at least as perfect as imperfect humans can create.''
& 7\%       & 858     & $-.92^{***}$                                                          \\\hline Library               & A source of extensive knowledge, like a library or encyclopedia                                                               &     ``AI is like a librarian that can recite information about every book on the shelves.''                                               
``AI is an electronic encyclopedia of all the existing information in the universe. When it’s asked a question, it can generate answers based on this encyclopedia.''& 5\%     & 610       & $.62^*$                                           \\\hline Future shaper           & A force that will shape the future in both positive and negative ways                                                         & ``A tech fad that will provide some actually useful uses in the coming years, but for now will mostly be used to advance data security violations for companies that already make too much money.''                                                                                                         ``I feel AI emerging is somewhat like when smartphones took over. It will show amazing capabilities but as a whole will make everyone lazier and less capable over time. It's as if you gave a caveman an object that creates fires and creates wheels without giving him the opportunity to learn for themselves.''                                                                                                                                                                                                                & 5\%    & 581        & $.39$ (ns)                                                         \\\hline Genie           & A mystical or mysterious being, such as a genie, wizard, or fortuneteller                                                     & ``The best metaphor is it being a Genie as it can create positive and negative things.''     ``I think the best metaphor for how AI works is like how a fortune teller can tell you about yourself. You can ask it questions about current situations or the future and it will provide you with information.''                                                                                                                                                                                                                                                                      & 4\%    & 498        & $.46$ (ns)                                \\\hline Mirror                & A mirror that reflects, imitates, or mimics humans                                                                            & ``It feels like a distorted mirror.''     ``AI is like Talking to an echo chamber. because So much of what it says is based off of what people have told it and what it has read.''                                                                                                                                                                                                                                                                                                  & 4\%       & 419     & $.69^{**}$                                                         \\\hline Child                 & A developing child that can learn and grow                                                                                    & ``AI is like a child that needs to learn, but is happy to provide output and is proud of it.''     ``AI works like a child. It absorbs and learns from those around it. It gathers information and eventually is able to do those things and create things on its own.''                                                                                                                                                                                                             & 4\%       & 419     & $.64^*$                                                              \\\hline Synthesizer  & A synthesizer that is able to combine elements creatively to form something new                                               & ``AI is like a painter.  You can give AI all the materials such as canvas, paint, paint brushes and a subject to draw.  With these materials, AI can put those things together and create something.''     ``AI is like an extension of my artistic abilities because through guidance I can get images close to what I would complete if I was using a camera and a subject.''                                                                                                      & 3\%        & 411    & $-.90^{***}$                       \\\hline Teacher               & A provider of knowledge and advice, like a teacher, professor, or mentor                                                      & ``AI is like a professor because it always has the answer.''    ``AI is like a coach because you can use it to guide and help in various topics.''                                                                                                                                                                                                                                                                              & 3\%        & 356    & $.75^{***}$                                        \\\hline Friend                & A friend with varying levels of reliability, closeness, and knowledge                                                         & ``AI is a like a friend you never had, but maybe never wanted.''     ``An AI works like a genius friend who you can ask pretty much anything but who doesn’t have a high emotional capacity.''                                                                                                                                                                                                                                                                                       & 2\%    & 286        & .76***                                           \\\hline Lifeform         & A lifeform that can self-propagate and grow, potentially beyond human control, such as plants and ecosystems                          & ``AI is a plant using all its resources given in its environment whether it be good or bad in order to produce an output (say a flower, fruit, etc).''
``AI is like a tree of knowledge because It is continuously growing and expanding intelligence''                                                                                                                                                                                                                                                                                                            & 2\%      & 229      & $.25$ (ns)                                 \\\hline Animal      & An animal, possibly serving as a companion (like a pet)                                                                           & ``AI works like training a dog how to do tricks, you show it how to do something and reward it for doing good things, then it learns and improves on doing those tricks'' ``AI is like a tame wolf because It is really cool but has potential to be dangerous''
& 2\%    & 210        & {$-.21$ (ns)}                   \\\hline Unexplored realm & A vast realm that remain largely unexplored and uncharted by humans, such as the ocean, jungles, and outer space                                & ``AI is like a jungle because it is vast and has interconnected network of pathways and holds a diverse variety of life forms'' ``AI is like the ocean, there is so much uncertainty and so much to discover''                                                                                                                                                                                                                                                                                         & 2\%   & 200         & {$.49$ (ns)}                         \\\hline God       & An always-there, god-like, omnipresent, and all-knowing presence.                                                                      & ``AI is like a God because It knows everything''     ``A deity or divine creature that has the knowledge and data to make informed decisions on a grand scale but lacks the consequences of decisions on an individual level. Basically, they are able to access more data than humans to make logical decisions but lack the empathy and morality that keep decisions fair for most.''                                                                                              & 1\%      & 141      & {$.66^*$}                                         \\\hline Folklore   & A character from folklore or a fairytale, such as the fairy godmother, Icarus, angel, satan/devil                               & ``AI is like a fairy godmother - there to help you with whatever you need''``AI is like a ghost because it's there and can help you but also might not be inherently good.''                                                                                                                                                                                                                                                                                                                                                  & 1\%  & 78          & {$.40$ (ns)}                                    \\\hline Thief                 & A thief of and threat to others' work and livelihoods                                                                                 & ``AI is like a thief because it uses others writings, drawings, and creativity to try to pass off as original. Also created to steal jobs.'' ``AI is like tracing over someone else’s art, maybe just changing colors, then calling it completely your own and not giving credit.''                                                                                                             & 0.5\%     & 57     & {$.25$ (ns)}  \\\hline                                
\end{tabular}
\caption{\textbf{20 dominant metaphors.} We identify 20 dominant metaphors (ordered by frequency (\%)) from the 12,000 metaphors collected using a combination of LM-based clustering and qualitative coding. $n$ is the count of the dominant metaphor, and $r$ is the Pearson's correlation between its frequency and time; positive/negative correlations indicate increases/decreases over time respectively. Additional examples are in Table \ref{tab:metaphor_examples}.}
\label{tab:metaphors}
\vspace{-3em}
\end{table*}

\begin{figure*}[ht]
    \vspace{-0.7em}
    \centering
    \includegraphics[width=0.28\linewidth]{ 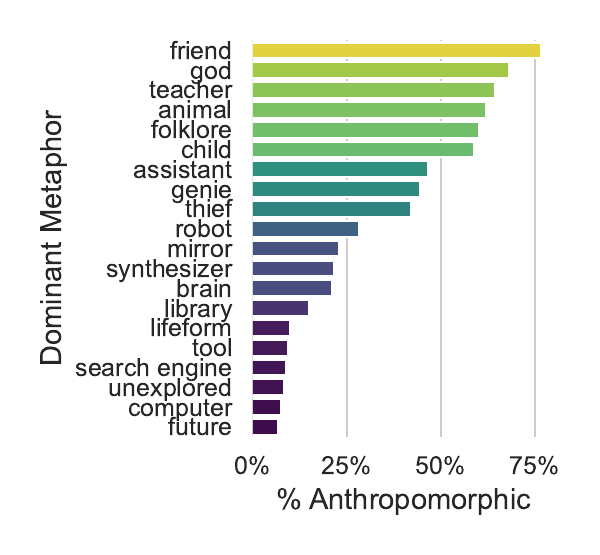}
    \includegraphics[width=0.7\linewidth]{ 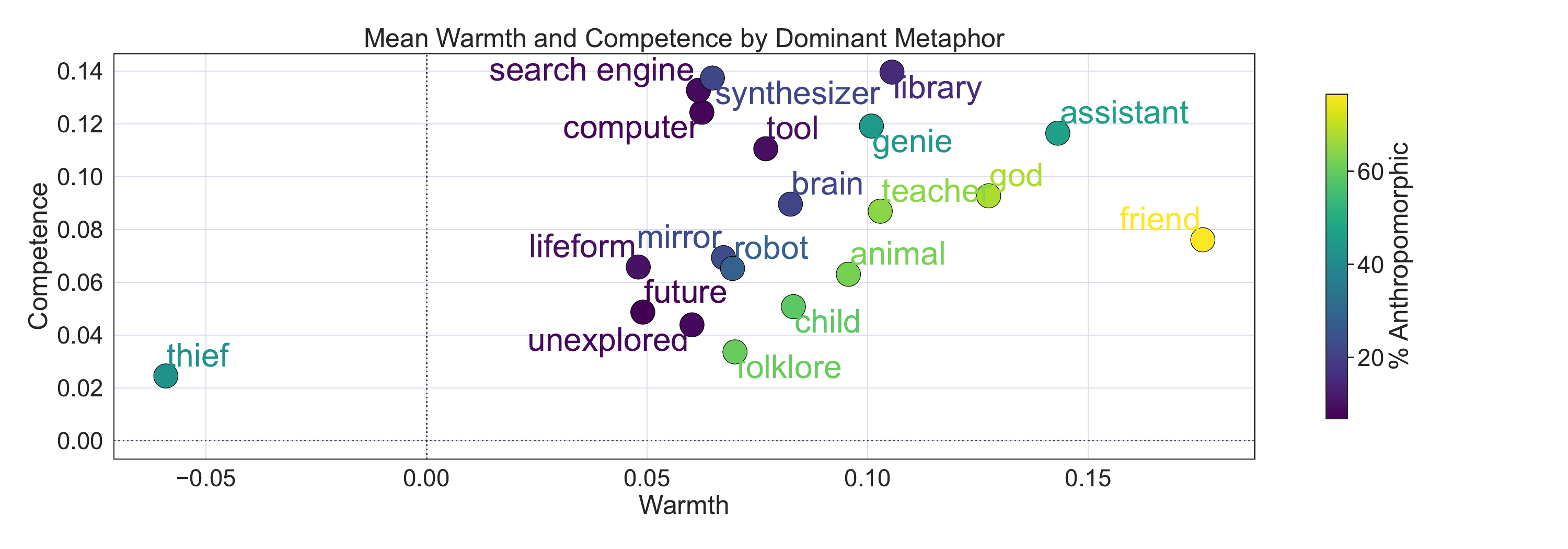}
    \caption{\textbf{Anthropomorphism percentage (left) and mean warmth and competence (right) for each dominant metaphor.} The dominant metaphors vary widely in anthropomorphism. They are all positive in both warmth and competence (except  ``thief'').
    }
    \label{fig:mean_by_cluster}
    \vspace{-0.7em}
\end{figure*}
\subsection{Implicit perceptions of AI as anthropomorphic, warm and competent over time}

The metaphors people use to describe AI can reveal the implicit perceptions they hold about AI. Applying our framework to measure implicit perceptions from the metaphors, we compute mean anthropomorphism, warmth, and competence scores for each dominant metaphor (Figure \ref{fig:mean_by_cluster}). While there is a wide range of anthropomorphism in dominant metaphors, we observe that every dominant metaphor has positive mean values of both warmth and competence (with the exception of the ``thief'' dominant metaphor, which has positive competence but negative mean warmth). Based on qualitative inspection of the individual metaphors, we find that this is because although some of the dominant metaphors 
reflect concepts that may have negative connotations, participants focus on warm and competent aspects of these concepts in their responses. For example, when describing AI as a ``child,'' participants often describe its learning and growing capabilities, and when describing AI as a ``genie,'' participants focus on its capabilities, which seem magical in their scope. For the dominant metaphor of ``unexplored realm'', rather than expressing fear of the unknown, participants' metaphors discuss the potential for discovery and capacity for ``wonder and beauty.'' Even for the dominant metaphor of ``thief,'' participants' responses describe, for instance, that AI is stealing work \textit{and} that it is able to accomplish many tasks through this stealing (more examples of individual metaphors are in Table \ref{tab:metaphor_examples}).

Beyond being warm and competent overall, we observe variation along these dimensions among the dominant metaphors. For example, the dominant metaphor of AI as a ``friend'' is higher in warmth but middling in competence, while the dominant metaphor of AI as ``search engine'' is associated with higher perceptions of competence but lower warmth. Note that perceiving AI as human-like does not necessarily engender positive perceptions; to the contrary, the anthropomorphic dominant metaphor of ``thief'' is lower in both warmth and competence, while the non-anthropomorphic dominant metaphor of AI as a ``library'' is higher in both. Nonetheless, when we measure the correlation between mean anthropomorphism and mean warmth and competence across the dominant metaphors (full details in Figure \ref{fig:correlations_2}), we find that mean anthropomorphism and warmth were significantly correlated (Pearson's $r = 0.45^*$). 
Indeed, dominant metaphors like friend, god, teacher, animal, assistant are all higher in anthropomorphism and warmth, while non-anthropomorphic dominant metaphors like tool or search engine are lower in warmth. We do not find a significant correlation between anthropomorphism and competence. At the individual metaphor level, this pattern persists but is much weaker: anthropomorphism is positively correlated with warmth ($r = 0.17^{***}$) and negatively correlated with competence ($r = -0.07^{***}$) (see full details in Figure \ref{fig:correlations_3}). This similarly suggest that anthropomorphism is more consistently linked to warmth than to competence.
\begin{figure*}[t]
    \centering
    \includegraphics[width=0.9\linewidth]{ 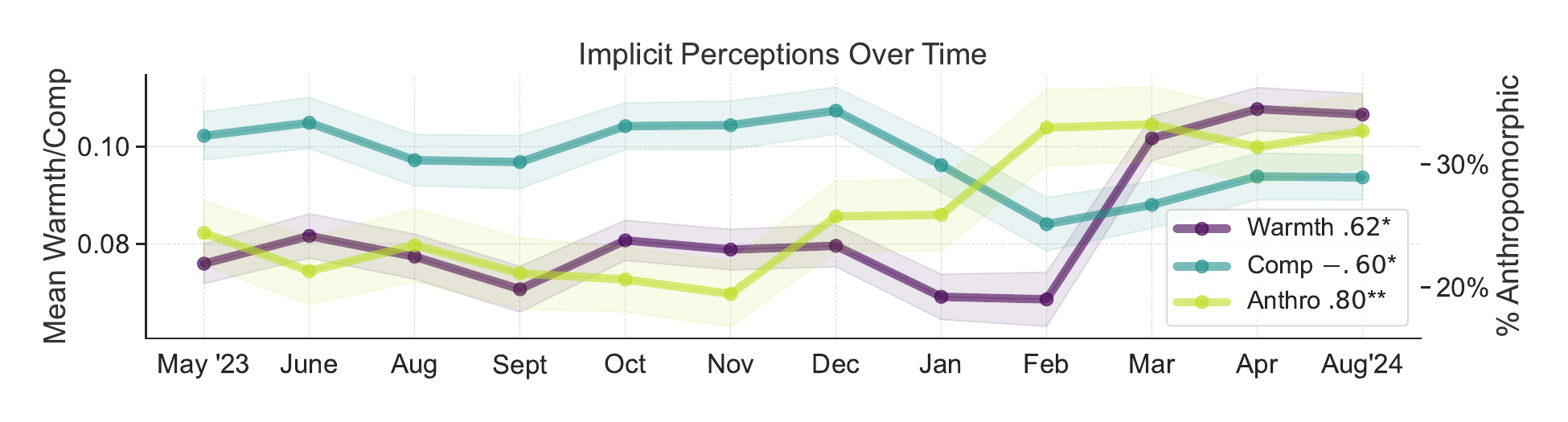}
    \includegraphics[trim={3cm 0 1cm 0},clip,width=0.9\linewidth]{ 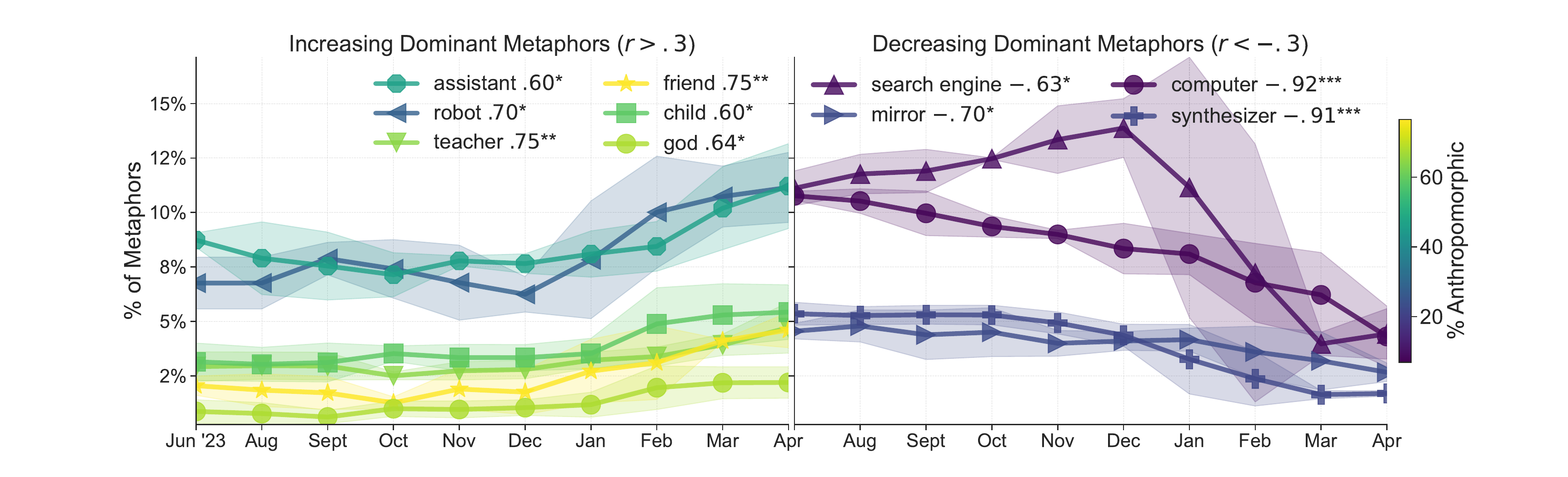}

    \caption{\textbf{Overall shifts in implicit perception scores over time (top)}. Shading reflects 95\% CI. For each month (x-axis), the left y-axis is mean warmth and competence, which are on a [-1, 1] range; the right y-axis is the percent of anthropomorphic metaphors by month. We find that anthropomorphic and warm metaphors are increasing in frequency over time, while competent metaphors are decreasing over time. \textbf{Dominant metaphors with statistically significant temporal change ($|r| > 0.3, p < 0.05$) (bottom)}. Each line represents the month-by-month prevalence of a dominant metaphor, color-coded based on its anthropomorphism percentage (the percent of metaphors in the cluster that are anthropomorphic). We find that anthropomorphic metaphors are increasing, while non-anthropomorphic metaphors are decreasing over time. Shading reflects 3-month rolling average. }
    \label{fig:topics_over_time}
        \vspace{-1em}

\end{figure*}

\paragraph{Temporal shifts: People view AI as increasingly human-like and warm}
Analyzing month-over-month shifts revealed how people’s perceptions changed over time. As shown in Figure \ref{fig:topics_over_time}, anthropomorphism increased by 34\% over the 12 months ($r = .80^{**}$). Understanding how dominant metaphors vary in prevalence over time provides insight into how different narratives about AI affect these perceptions: people became more likely to describe AI as being like a teacher ($r = .75^{***}$), a friend ($r = .75^{**}$), or an assistant ($r = .60^*$), and less likely to see it as a distinctly non-human entity like a computer ($r = -.92^{***}$), a search engine ($r = -.63^*$), or a mirror ($r = -.70^*$) (Figure \ref{fig:topics_over_time}).
In addition to seeing AI as more human-like, implicit perceptions of AI as warm increased by 41\% over time ($r = .62^*$). Notably, however, this did not correspond to an increased perception that AI is competent. Instead, implicit perceptions of AI as competent decreased by 8\% over time ($r = -.60^*$). This may corroborate previous work finding that people broadly prefer AI that is conceptualized with metaphors that signal relatively lower competence \cite{khadpe2020conceptual}. In particular, the highest-warmth dominant metaphors describing AI as a friend, assistant, god ($r = .64^*$), and teacher became increasingly common over time, making up four of the six dominant metaphors with statistically significant temporal increase in prevalence. (The other increasing dominant metaphors, ``child'' ($r = .60^*$, 8th warmest) and ``robot'' ($r = .70^*$, 12th warmest), also rank relatively high in warmth.) In contrast, three of the top four highest-competence dominant metaphors—computer, synthesizer ($r = -.91^{***}$), and search engine—are the three most significantly decreasing clusters. Taken together, our findings reflect a societal shift in seeing AI as being more human-like and warm. We situate these changes in the context of rapid sociotechnical developments during this time period, such as the rapid increase in use of AI assistants \cite{gabriel2024ethics} and the release of major user-facing AI models like OpenAI's GPT-4 \cite{achiam2023gpt}, Meta's Llama \cite{touvron2023llama}, and Google’s Gemini \cite{team2023gemini}.
\begin{figure*}
    \centering
    \includegraphics[width=0.52\linewidth]{ 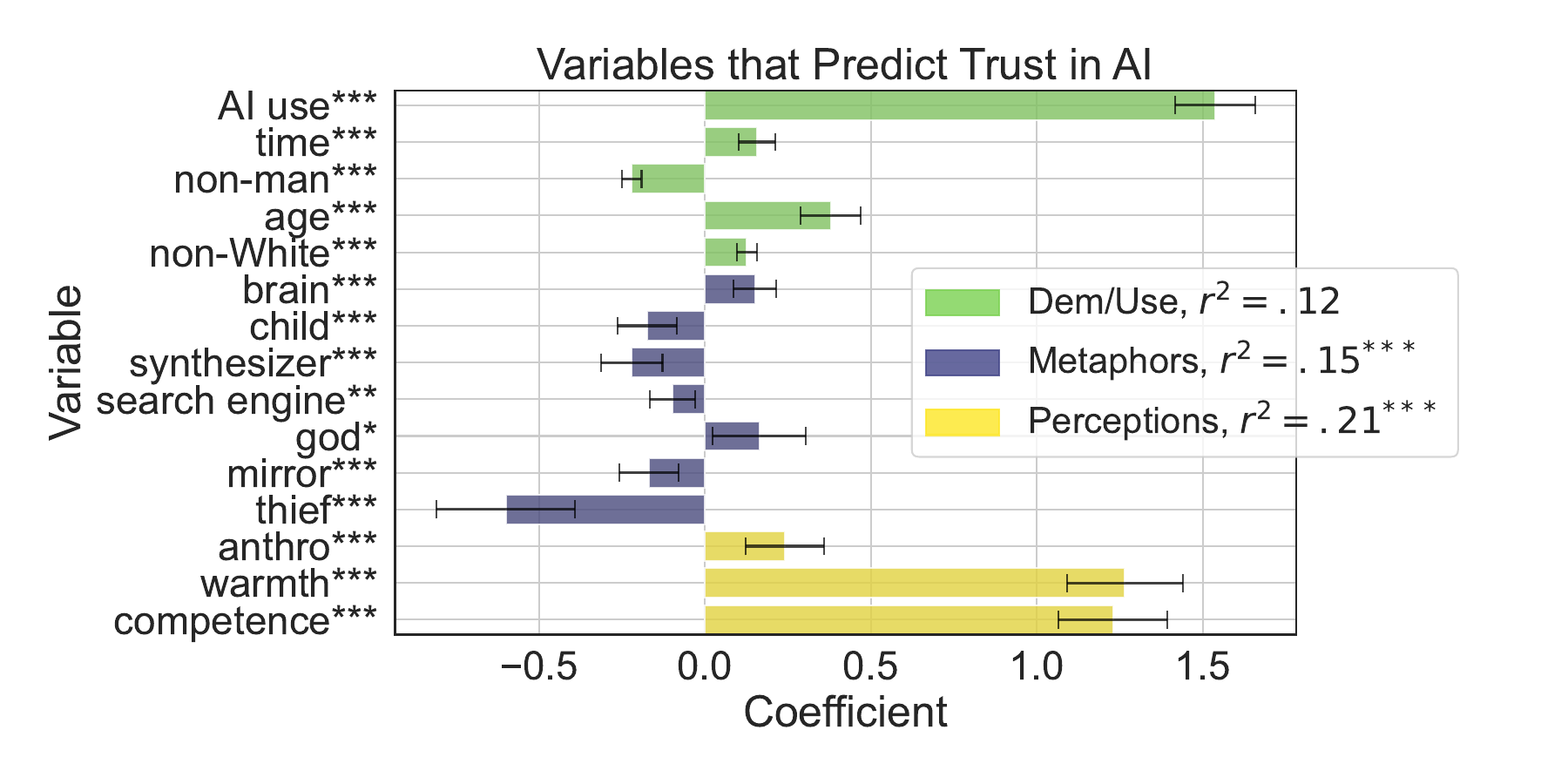}
\includegraphics[width=0.47\linewidth]{ 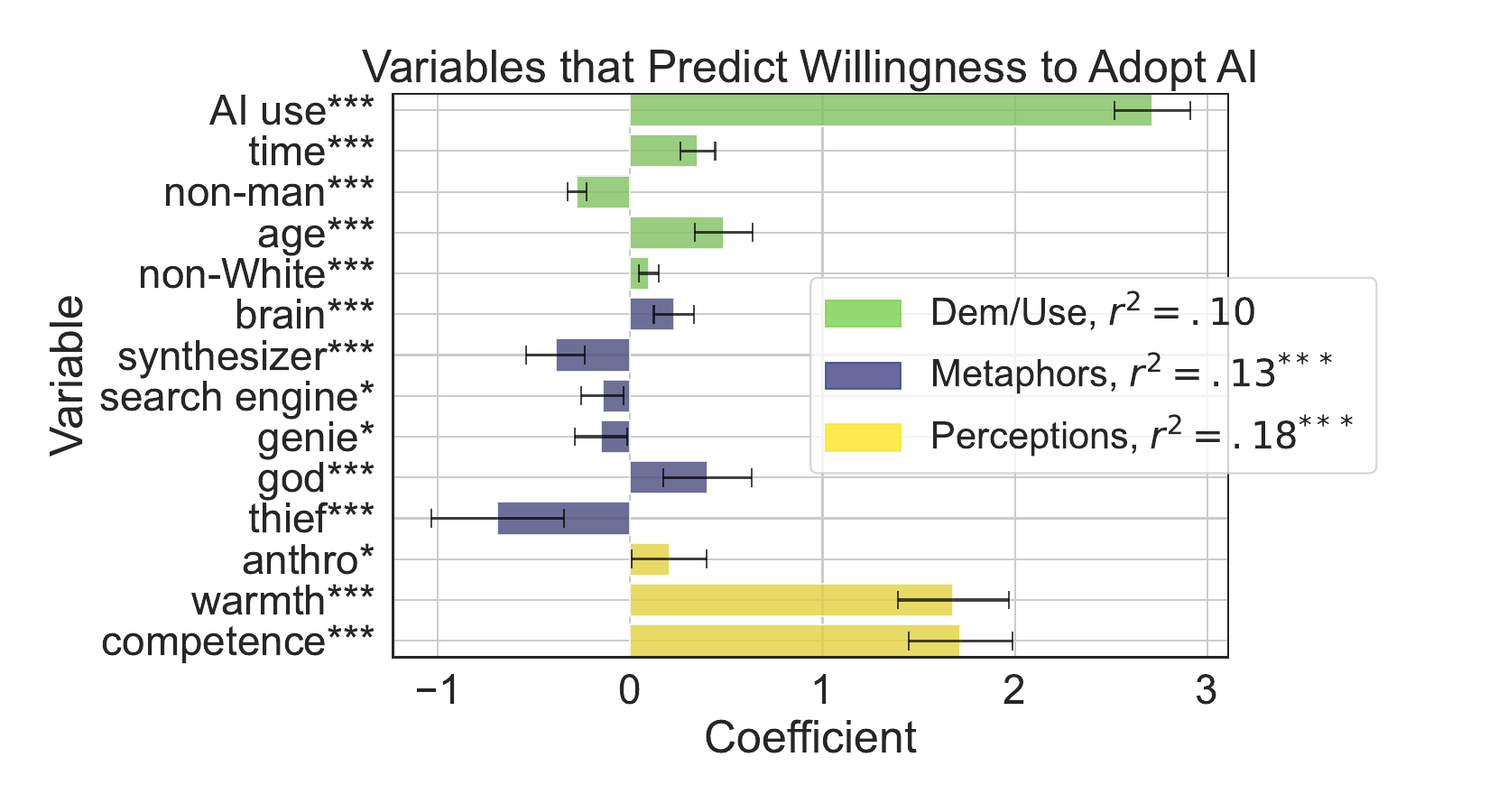}
    \caption{\textbf{Statistically significant variables in regression combining all three blocks: demographics and use (green), dominant metaphors (purple), and implicit perceptions (yellow).} Frequency of AI use is the largest predictor for both trust and willingness to adopt AI, followed by warmth and competence, which are comparable to frequency of AI use for predicting trust.}
    \label{fig:3blocks}
\end{figure*}
\subsection{Understanding attitudes: Dominant metaphors and implicit perceptions explain trust and adoption}
In our hierarchical regression to predict attitudes from metaphors and implicit perceptions, the first block of variables included demographic and AI usage variables, which explain $12\%$ and $10\%$ of the variance in trust and willingness to adopt AI, respectively (adjusted $r^2 = 0.12, 0.10$). The second block of variables, which included the dominant metaphors, explain $3\%^{***}$ and $2\%^{***}$ more variance for  trust and adoption of AI respectively. Finally, the third block of variables, implicit perceptions, help explain an additional $6\%^{***}$ and $5\%^{***}$ of variance ( adjusted $r^2 = 0.21, 0.18$). Together, the dominant metaphors and implicit perceptions explain $75\% $ and $80\%$ more of the variance than demographics and frequency of use alone. (See detailed results in Table \ref{tab:regressioncoef}).
\paragraph{Specific metaphors and perceptions associated with trust and adoption} 
Regression coefficients from the full model (Figure \ref{fig:3blocks}) show that, beyond AI use, warmth and competence are the strongest predictors of trust and AI adoption, with effects similar in magnitude to AI use. This is likely due to warm and competent metaphors such as ``assistant,'' ``friend,'' ``teacher,'' and ``library,'' which stand out as predicting trust and adoption positively in a dominant-metaphor-only regression (though these effects appear mediated by the warmth and competence variables in the full model -- see Appendix Fig. \ref{fig:regressions}). Anthropomorphism also predicts positive trust and adoption, but to a lesser extent.
Notably, even in combination with the perceptual variables, ``thief'' strongly predicts low trust and reluctance to adopt, while ``brain'' and ``god'' are linked to higher trust and adoption. 
Building on previous work on the nuanced relationship between anthropomorphism and trust \cite{zhou2024rel,cohn2024believing}, we find specific metaphors that reveal what topics are salient to people's trust and adoption: conceptualizations of AI as a helpful human-like entity (``friend'', ``teacher'', ``assistant''), or as a source of vast amounts of knowledge  (``library'', ``god'', ``brain'') facilitate trust and adoption. In contrast, anthropomorphic conceptualizations of AI as a human-like ``thief'' are often related to the widespread public concerns about copyright issues and AI plagiarizing or stealing creative work \cite{goetze2024ai,samuelson2023generative}. 

\subsection{Demographic differences reveal disparities in trust and adoption}

\begin{figure*}
    \centering
\includegraphics[width=0.32\linewidth]{ 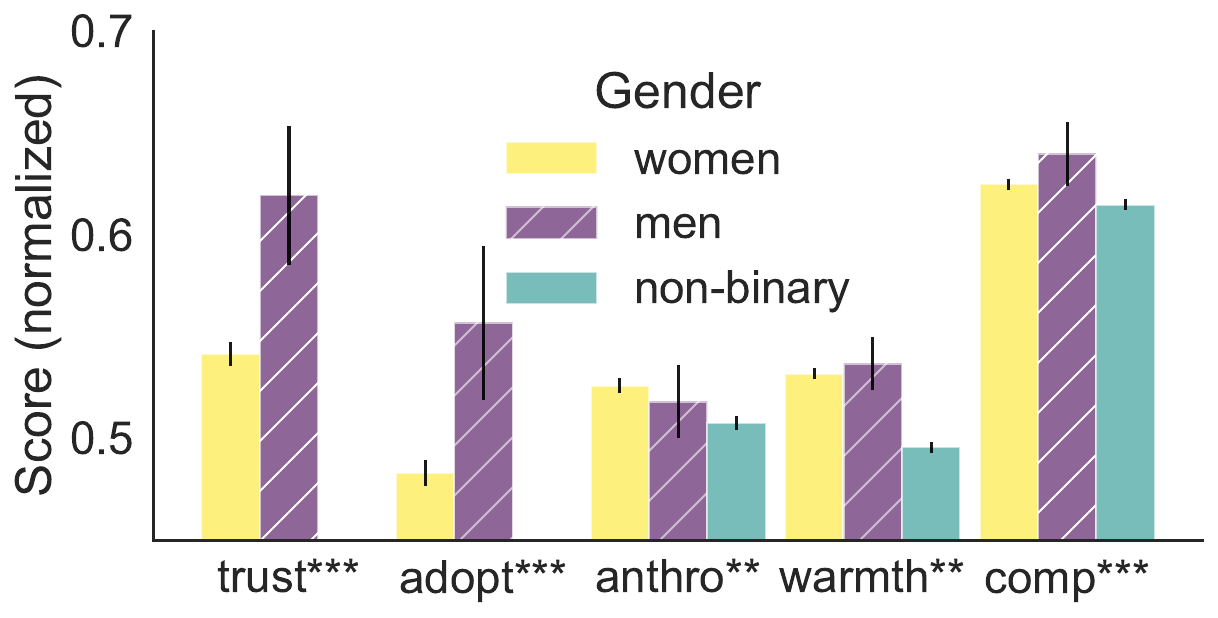}
\includegraphics[width=0.32\linewidth]{ 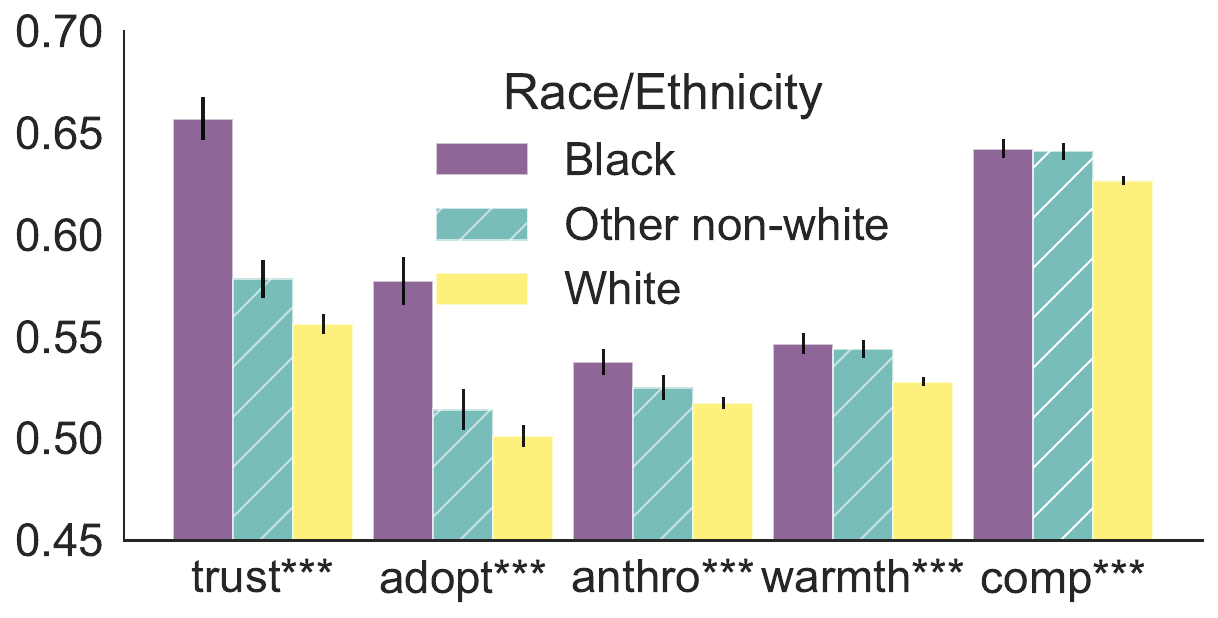}
\includegraphics[width=0.32\linewidth]{ 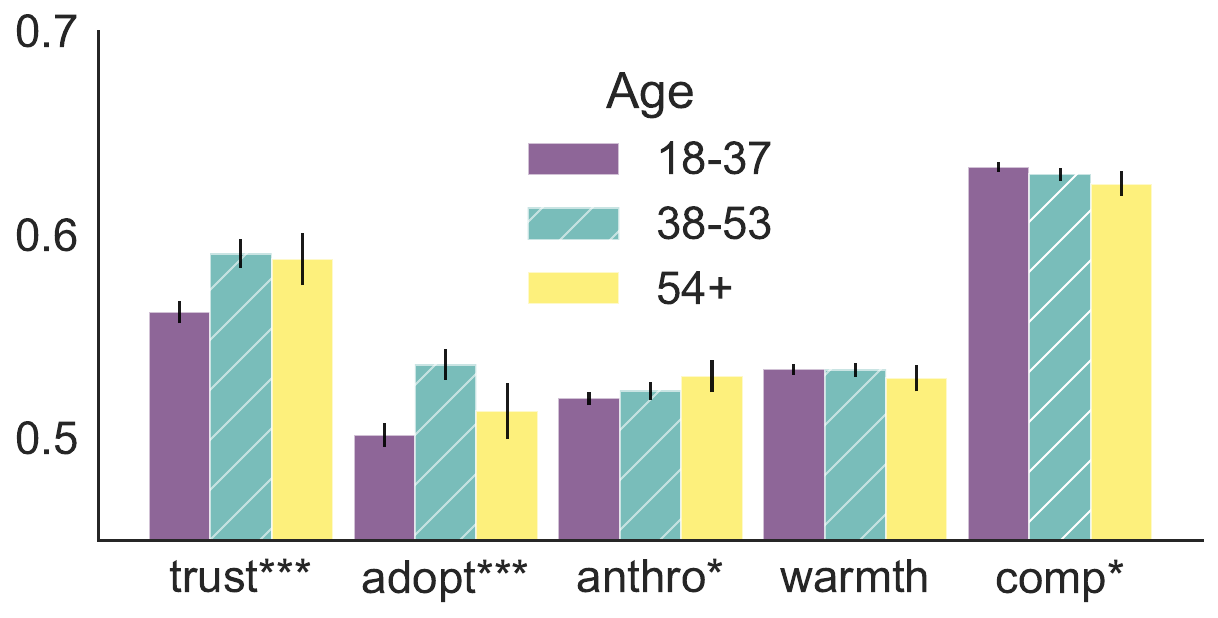}
    \caption{\textbf{Demographic differences in implicit perceptions and attitudes.}  Implicit perceptions labeled with *s have statistically significant demographic differences based on a two-sample t-test (binarized as between men and non-men for gender; between white and non-white for race/ethnicity; between 18-37 and 37+ for age). All scores are normalized to 0-1 to be comparable. Error bars represent 95\% CI. We find that men trust and are more willing to adopt AI, and they view it as more competent and warm but less anthropomorphic. We find that non-white participants trust and are more likely to adopt AI, and they have more anthropomorphic, warm, and competent metaphors of AI. We also identify that older participants trust AI more and anthropomorphize it more. More detailed breakdowns are in Appendix \ref{sec:moredetail}.} 
    \label{fig:dem_diff}

\end{figure*}

Corroborating prior work that women have more negative attitudes toward AI \cite{zhang2019artificial,pewresearchcenter_2024_trust}, we find that women are less trusting of and less willing to adopt AI. However, in contrast to previous work finding that race is not a significant predictor of AI attitudes and that trust and age are negatively correlated \cite{zhang2019artificial}, our findings reveal unexpected patterns in AI trust and adoption: \textit{surprisingly, non-white and older people trust AI more and are more willing to adopt AI}, i.e., older participants and non-white participants (and in particular Black participants) reported significantly higher levels of trust in AI and greater willingness to adopt these technologies compared to younger and white participants respectively (Figure \ref{fig:dem_diff}). Dominant metaphors and implicit perceptions provide insights into these phenomena, though additional research is needed to establish causal relationships. Full details of demographic differences in dominant metaphors are in Figure \ref{fig:dem_diff_metaphors}.

\paragraph{Gender: Men view AI as warmer and more competent, but less human-like.}
Our survey corroborates previous work that women overall view AI in more negative light and are significantly less trusting and willing to adopt AI compared to men \cite{zhang2019artificial,pewresearchcenter_2024_trust}. Women and non-binary people's metaphors are overall less warm and competent, though they are also significantly more anthropomorphic. 
Differences in rates of dominant metaphors give some insight into these patterns: based on a two-sample binomial test, ``robot'' and ``genie'' are disproportionately more frequently provided by women and non-binary people. In contrast, men significantly more frequently provide highly competent or warm--but not necessarily anthropomorphic--metaphors like ``teacher'', ``search engine'', and ``computer'', invoking entities that are familiar and helpful. 
Note that this may also reflect that women tend to anthropomorphize entities more due to social and cultural factors \cite{chin2004measuring} and reflect broader gendered differences in expression \cite{lakoff1973language,baron2002extreme}.
\paragraph{Race/Ethnicity: Participants of color view AI as more warm, competent, and anthropomorphic.}
Non-white participants, who surprisingly trust AI more and are more willing to adopt it, also provide metaphors that reflected higher levels of warmth, competence, and anthropomorphism. This pattern emerges despite well-documented concerns about AI systems performing poorly for non-white users \cite{buolamwini2018gender,hoffmann2019fairness} and perpetuating harmful racial stereotypes \cite{weidinger2022taxonomy,cheng-etal-2023-marked,bianchi2023easily}. Non-white participants more frequently described AI through metaphors suggesting power and agency: based on a two-sample
binomial test, the ``genie'' and ``god'' dominant metaphors--evoking magic and mysticism--were significantly more
common among non-white participants. In contrast, white participants more frequently used the ``search engine''
metaphor, which conveys a more limited and controllable entity.

\paragraph{Age: Older people trust AI more and view it as more anthropomorphic but less competent.}
Using a two-sample binomial test, we find that the dominant metaphor of ``friend'' is significantly more common among those older than 38, while dominant metaphors of ``search engine'', ``genie'', ``mirror'', ``animal'', and ``child'' are all significantly more common among the 18-37 age groups. This perhaps suggests that older people may be more curious about AI companions and see that as the primary potential use of AI, though further research is needed to understand this phenomenon.
\section{Discussion}

\subsection{Navigating appropriate levels of trust and adoption}

Our findings provide evidence that certain framings and conceptualizations of AI are particularly risky and require more care to deploy because they project warmth and human-likeness, which may lead to unwarranted trust and overreliance. For instance, our work shows that AI-based technologies that are conceptualized as a ``friend'' or ``assistant'' are associated with higher levels of trust, which may not match the reality of a system that is prone to generating false information \cite{sahoo-etal-2024-comprehensive}, fundamentally unable to make sincere statements \cite{kasirzadeh2023conversation}, or poses other risks \cite{manzini2024code}. 
Moreover, our finding that anthropomorphic and warm perceptions (which facilitate trust and adoption) have rapidly increased suggest the urgent need to conceptualize AI in ways that lead to more appropriate levels of trust and adoption. 
Rather than thinking of AI as a ``genie'' or ``god'', it may be more accurate to conceptualize it as a ``tool'' to use. AI practitioners, designers, and leaders should be mindful of the consequences of the figurative language they use, which may inadvertently impact users' behavior toward and with AI. Our findings also underscore the importance of AI literacy  \cite{ng2021conceptualizing}, of which accurate conceptualizations of AI is a key component \cite{roe2024funhouse,gupta2024assistant}. 

Beyond the metaphors we collected, our framework for measuring anthropomorphism, warmth, and competence from metaphors is useful to understand how the public will respond to new conceptualizations beyond those currently dominant in the public imaginary. By leveraging metaphors that align with desired levels of of anthropomorphism, warmth and competence as measured by our framework, designers can invoke conceptualizations of AI that more accurately reflect their capabilities and use cases.

\subsection{Using metaphors and implicit perceptions to address inequities in AI adoption}
Our demographic analyses reveal that women and non-binary people, non-white participants, and older participants all provide more anthropomorphic metaphors of AI. This suggests that the negative consequences of anthropomorphic AI, including overreliance \cite{akbulut2024all}, deception \cite{winkle2021assessing}, dehumanization \cite{Bender2024-de}, and stereotyping \cite{abercrombie-etal-2023-mirages} are a higher risk for these populations. This reifies existing digital inequalities and ways that marginalized populations are disproportionately harmed by technology \cite{robinson2015digital, lutz2019digital,hofmann2024ai,mohamed2020decolonial}. Moreover, we notice that the dominant metaphor of ``genie'' are more common among women and non-white participants, while more functional dominant metaphors like ``search engine'', ``synthesizer'', and ``computer'' are more common among white and male participants. The notion of a ``genie'' connotes the expectation that AI has mystical or magical capabilities, which again raises concerns of overreliance particularly for these populations. The power differential between these top dominant metaphors may also reflect different perspectives on personal agency and control in relation to AI systems.  Building on related work evaluating justice across AI applications \cite{ajmani2024data} and challenging existing power differentials exacerbated by technology \cite{d2023data}, we point to dominant metaphors and implicit perceptions not only as a way to reveal inequities via differences in public attitudes, but also as a starting point to consider how conceptualizations may help to mediate inequities. Reliable, safe, and trustworthy AI should enable a high amount of human control: like technologies like elevators and cameras that are highly automated but have a high amount of human control \cite{shneiderman2022human}, AI must be built in a way that enables all people to conceptualize it as a tool that can be useful to them \cite{birhane2022power}, rather than a force that threatens their agency and humanness ~\cite{aizenberg2020designing,van2024artificial}. 

\section{Conclusion and Future Work}

Our study examines the evolution of public perceptions of AI over time among the American public through a novel dataset of 12,000 metaphors. Within the short timespan of just a year and a half, we see substantial shifts in how people think and feel about AI -- highlighting the impact that the rapid adoption of systems like ChatGPT and Gemini have on our collective consciousness.
First, the identification of dominant metaphors, ranging from tools and teachers to pets and thieves, underscores the current complexity of public perceptions. These metaphors reveal not only how people understand and relate to AI, but also how they rationalize its capabilities and limitations. The increasing prevalence of anthropomorphic and warm metaphors, such as those likening AI to teachers or assistants, signals a growing tendency to ascribe social and human-like qualities to AI systems. These shifts have significant implications: as AI is increasingly integrated into everyday life, and perceived as more human-like and warm, people may be more inclined to trust and adopt these technologies across various domains. 
However, this also raises critical ethical questions about over-reliance and misplaced trust in systems that, despite their perceived warmth, remain fundamentally non-human and limited in their capabilities \cite{porra2020can,kasirzadeh2023conversation}. In particular, marginalized groups may be more susceptible to these potential risks since they have more anthropomorphic perceptions of AI. More broadly, our methodology of capturing metaphors and measuring implicit perceptions enable understanding public perceptions of AI at a large scale, tracking longitudinal changes, anticipating future shifts in public attitudes, and providing actionable insights for AI stakeholders.

\paragraph{Understanding causality} Our data leverages month-over-month data from nationally representative samples to understand shifts in the American public's perceptions of AI. However, the cross-sectional nature of our data precludes us from drawing causal inferences regarding the effects of personal and public AI adoption on people's metaphors, or the effects of people's metaphors of AI on their behavior. Rather, our data provide a window into how everyday Americans understand these complex sociotechnical systems during a period of significant public interest and uptake of AI (May 2023 to August 2024). Our method allows us to examine the broad strokes of shifts in public perceptions of AI by leveraging large-scale data, a scalable analytical framework, and a prompt that enables people even with low literacy around AI to share how they feel. Going forward, future research may expand on this work by linking social and cultural changes (i.e., the deployment of a new widely used tool, the implementation of a new policy regulating AI use) to changes in the dominant metaphors people have about AI. Furthermore, scholars may complement our month-by-month data by conducting longitudinal within-person analyses following the same cohort of individuals to examine if and how interacting with AI influences their metaphors, perceptions, and behaviors. 

\paragraph{Looking beyond the American context} 
While our approach allowed us to obtain granular insight into how Americans of different ages, genders, and ethnicities thought about AI and responded to this sea-change in the communication landscape, this focus on one country is a key limitation to our work. Research on the effects of sociotechnical systems has disproportionately focused on people in Western, and particularly American, contexts \cite{ghai2021s}. Additional research is needed to contextualize the metaphors people have about AI in different countries and
cultures across both the Global North and Global South \cite{kelley2021exciting}.

\section{Limitations}
Our survey design had several limitations. First, we collected usage data only on AI-based chatbots, and we did not collect data on more implicit usages of AI such as AI in search or text completion. Additionally, the survey design may have introduced priming effects, as participants first answered the open-ended metaphor question before responding to the trust and adoption questions. However, we believe these limitations do not undermine the validity of our findings.

Moreover, there may be other important dimensions of AI perception beyond anthropomorphism, warmth, and competence that were not captured in our study. Our contribution is novel in that it offers a framework to quantify these dimensions from open-ended text, and we encourage future work to extend and refine this framework.

\bibliographystyle{ACM-Reference-Format}
\bibliography{biblio,anthro}

\appendix

\renewcommand{\thetable}{A\arabic{table}}

\renewcommand{\thefigure}{A\arabic{figure}}

\setcounter{figure}{0}

\setcounter{table}{0}

\begin{table}[]\tiny
\begin{tabular}{llllll}
Month     & Total & Gender                                                                                                             & Ethnicity                                                                                                                           & Age                                                                                                                                              & Education                                                                                                                                                          \\
May       & 1,182 & \begin{tabular}[c]{@{}l@{}}Men: 49.6\%\\    \\ Women: 49.3\%\\    \\ Non-binary/\\    \\ other: 1.1\%\end{tabular} & \begin{tabular}[c]{@{}l@{}}White: 73.8\%\\    Black: 11.4\%\\    \\ Asian: 6.6\%\\    Latino: 5.2\%\\    Other: 3\%\end{tabular}    & \begin{tabular}[c]{@{}l@{}}18-24: 5.1\%\\    25-34: 32\%\\    35-44: 28.6\%\\    45-54: 19.6\%\\    55-64: 12\%\\    65+: 2.7\%\end{tabular}     & \begin{tabular}[c]{@{}l@{}}Some high   school: .4\%\\    High school: 10.4\%\\    Some college: 12.1\%\\    College: 53.6\%\\    Postgraduate: 23.5\%\end{tabular} \\
June      & 1,008 & \begin{tabular}[c]{@{}l@{}}Men: 49.6\%\\    \\ Women: 48.8\%\\    \\ Non-binary/\\    \\ other: 1.6\%\end{tabular} & \begin{tabular}[c]{@{}l@{}}White: 71.2\%\\    Black: 13.3\%\\    \\ Asian: 4.8\%\\    Latino: 8\%\\    Other: 2.7\%\end{tabular}    & \begin{tabular}[c]{@{}l@{}}18-24: 9.6\%\\    25-34: 36.5\%\\    35-44: 24.3\%\\    45-54: 18.6\%\\    55-64: 8.8\%\\    65+: 2.2\%\end{tabular}  & \begin{tabular}[c]{@{}l@{}}Some high   school: .4\%\\    High school: 9.3\%\\    Some college: 9.4\%\\    College: 54.1\%\\    Postgraduate: 26.9\%\end{tabular}   \\
July      & 846   & \begin{tabular}[c]{@{}l@{}}Men: 55.1\%\\    \\ Women: 43.4\%\\    \\ Non-binary/\\    \\ other: 1.5\%\end{tabular} & \begin{tabular}[c]{@{}l@{}}White: 71.5\%\\    Black: 10.8\%\\    \\ Asian: 6.3\%\\    Latino: 8.\%\\    Other: 3.2\%\end{tabular}   & \begin{tabular}[c]{@{}l@{}}18-24: 10.2\%\\    25-34: 38.5\%\\    35-44: 25.9\%\\    45-54: 15.5\%\\    55-64: 8.6\%\\    65+: 1.3\%\end{tabular} & \begin{tabular}[c]{@{}l@{}}Some high   school: .3\%\\    High school: 11.5\%\\    Some college: 11.8\%\\    College: 49.9\%\\    Postgraduate: 26.5\%\end{tabular} \\
August    & 928   & \begin{tabular}[c]{@{}l@{}}Men: 51.3\%\\    \\ Women: 45.9\%\\    \\ Non-binary/\\    \\ other: 2.8\%\end{tabular} & \begin{tabular}[c]{@{}l@{}}White: 74.7\%\\    Black: 9.9\%\\    \\ Asian: 7.3\%\\    Latino: 5.9\%\\    Other: 2.2\%\end{tabular}   & \begin{tabular}[c]{@{}l@{}}18-24: 9.8\%\\    25-34: 41.4\%\\    35-44: 25.9\%\\    45-54: 14.9\%\\    55-64: 6.8\%\\    65+: 1.4\%\end{tabular}  & \begin{tabular}[c]{@{}l@{}}Some high   school: .8\%\\    High school: 10.8\%\\    Some college: 11\%\\    College: 49.9\%\\    Postgraduate: 27.5\%\end{tabular}   \\
September & 878   & \begin{tabular}[c]{@{}l@{}}Men: 41.2\%\\    \\ Women: 56.9\%\\    \\ Non-binary/\\    \\ other: 1.8\%\end{tabular} & \begin{tabular}[c]{@{}l@{}}White: 72.3\%\\    Black: 9.2\%\\    \\ Asian: 7.4\%\\    Latino: 7.4\%\\    Other: 3.6\%\end{tabular}   & \begin{tabular}[c]{@{}l@{}}18-24: 13.9\%\\    25-34: 43.4\%\\    35-44: 24.8\%\\    45-54: 11.8\%\\    55-64: 5.7\%\\    65+: .5\%\end{tabular}  & \begin{tabular}[c]{@{}l@{}}Some high   school: .4\%\\    High school: 9.3\%\\    Some college: 9.4\%\\    College: 54.1\%\\    Postgraduate: 26.9\%\end{tabular}   \\
October   & 1,020 & \begin{tabular}[c]{@{}l@{}}Men: 42.4\%\\    \\ Women: 56.3\%\\    \\ Non-binary/\\    \\ other: 1.4\%\end{tabular} & \begin{tabular}[c]{@{}l@{}}White: 67.8\%\\    Black: 10.8\%\\    \\ Asian: 9.1\%\\    Latino: 8.3\%\\    Other: 3.9\%\end{tabular}  & \begin{tabular}[c]{@{}l@{}}18-24: 10.4\%\\    25-34: 39.9\%\\    35-44: 24.9\%\\    45-54: 15.8\%\\    55-64: 8.3\%\\    65+: .7\%\end{tabular}  & \begin{tabular}[c]{@{}l@{}}Some high   school: .1\%\\    High school: 8.9\%\\    Some college: 10.9\%\\    College: 49.7\%\\    Postgraduate: 30.4\%\end{tabular}  \\
November  & 965   & \begin{tabular}[c]{@{}l@{}}Men: 45.1\%\\    \\ Women: 52.8\%\\    \\ Non-binary/\\    \\ other: 2.1\%\end{tabular} & \begin{tabular}[c]{@{}l@{}}White: 59.5\%\\    Black: 18.3\%\\    \\ Asian: 9.8\%\\    Latino: 8.0\%\\    Other: 4.4\%\end{tabular}  & \begin{tabular}[c]{@{}l@{}}18-24: 8.6\%\\    25-34: 39.7\%\\    35-44: 27.5\%\\    45-54: 16.3\%\\    55-64: 6.3\%\\    65+: 1.7\%\end{tabular}  & \begin{tabular}[c]{@{}l@{}}Some high   school: .5\%\\    High school: 8.9\%\\    Some college: 12.3\%\\    College: 49.8\%\\    Postgraduate: 28.6\%\end{tabular}  \\
December  & 998   & \begin{tabular}[c]{@{}l@{}}Men: 46\%\\    \\ Women: 51.5\%\\    \\ Non-binary/\\    \\ other: 2.5\%\end{tabular}   & \begin{tabular}[c]{@{}l@{}}White: 62.8\%\\    Black: 14.4\%\\    \\ Asian: 10.9\%\\    Latino: 7.7\%\\    Other: 4.1\%\end{tabular} & \begin{tabular}[c]{@{}l@{}}18-24: 11.3\%\\    25-34: 39.3\%\\    35-44: 26\%\\    45-54: 15.2\%\\    55-64: 6.4\%\\    65+: 1.8\%\end{tabular}   & \begin{tabular}[c]{@{}l@{}}Some high   school: .5\%\\    High school: 6\%\\    Some college: 9.3\%\\    College: 53.8\%\\    Postgraduate: 30.5\%\end{tabular}     \\
January   & 1,003 & \begin{tabular}[c]{@{}l@{}}Men: 38.1\%\\    \\ Women: 59.9\%\\    \\ Non-binary/\\    \\ other: 2\%\end{tabular}   & \begin{tabular}[c]{@{}l@{}}White: 63.6\%\\    Black: 16.4\%\\    \\ Asian: 8.9\%\\    Latino: 7.3\%\\    Other: 3.9\%\end{tabular}  & \begin{tabular}[c]{@{}l@{}}18-24: 11.9\%\\    25-34: 42.3\%\\    35-44: 25.3\%\\    45-54: 12.8\%\\    55-64: 6.5\%\\    65+: 1.3\%\end{tabular} & \begin{tabular}[c]{@{}l@{}}Some high   school: .2\%\\    High school: 6.8\%\\    Some college: 8.7\%\\    College: 53.4\%\\    Postgraduate: 30.8\%\end{tabular}   \\
February  & 1,040 & \begin{tabular}[c]{@{}l@{}}Men: 41.4\%\\    \\ Women: 57.5\%\\    \\ Non-binary/\\    \\ other: 1.1\%\end{tabular} & \begin{tabular}[c]{@{}l@{}}White: 56.2\%\\    Black: 21.3\%\\    \\ Asian: 13.9\%\\    Latino: 5.7\%\\    Other: 2.8\%\end{tabular} & \begin{tabular}[c]{@{}l@{}}18-24: 7.4\%\\    25-34: 33.1\%\\    35-44: 26.3\%\\    45-54: 20.7\%\\    55-64: 10.6\%\\    65+: 1.9\%\end{tabular} & \begin{tabular}[c]{@{}l@{}}Some high   school: .1\%\\    High school: 8.4\%\\    Some college: 11.6\%\\    College: 52\%\\    Postgraduate: 28\%\end{tabular}      \\
March     & 1,025 & \begin{tabular}[c]{@{}l@{}}Men: 45\%\\    \\ Women: 52.9\%\\    \\ Non-binary/\\    \\ other: 2.1\%\end{tabular}   & \begin{tabular}[c]{@{}l@{}}White: 63.6\%\\    Black: 16.2\%\\    \\ Asian: 8.8\%\\    Latino: 7.3\%\\    Other: 4.1\%\end{tabular}  & \begin{tabular}[c]{@{}l@{}}18-24: 10.9\%\\    25-34: 38.2\%\\    35-44: 27.8\%\\    45-54: 15.3\%\\    55-64: 6.5\%\\    65+: 1.2\%\end{tabular} & \begin{tabular}[c]{@{}l@{}}Some high   school: .3\%\\    High school: 7.7\%\\    Some college: 9.1\%\\    College: 54.6\%\\    Postgraduate: 28.3\%\end{tabular}   \\
April     & 1,018 & \begin{tabular}[c]{@{}l@{}}Men: 47.3\%\\    \\ Women: 51.3\%\\    \\ Non-binary/\\    \\ other: 1.4\%\end{tabular} & \begin{tabular}[c]{@{}l@{}}White: 61.1\%\\    Black: 18.0\%\\    \\ Asian: 9.9\%\\    Latino: 8.3\%\\    Other: 2.7\%\end{tabular}  & \begin{tabular}[c]{@{}l@{}}18-24: 8.5\%\\    25-34: 33.4\%\\    35-44: 29.5\%\\    45-54: 20.3\%\\    55-64: 6.5\%\\    65+: 1.8\%\end{tabular}  & \begin{tabular}[c]{@{}l@{}}Some high   school: .3\%\\    High school: 7.5\%\\    Some college: 11.2\%\\    College: 52.4\%\\    Postgraduate: 28.6\%\end{tabular}  \\
August    & 992   & \begin{tabular}[c]{@{}l@{}}Men: 46.1\%\\    \\ Women: 52.3\%\\    \\ Non-binary/\\    \\ other: 1.6\%\end{tabular} & \begin{tabular}[c]{@{}l@{}}White: 62.8\%\\    Black: 14.0\%\\    \\ Asian: 9.6\%\\    Latino: 10.2\%\\    Other: 3.4\%\end{tabular} & \begin{tabular}[c]{@{}l@{}}18-24: 4.5\%\\    25-34: 35.4\%\\    35-44: 32.3\%\\    45-54: 20.4\%\\    55-64: 6.1\%\\    65+: 1.3\%\end{tabular}  & \begin{tabular}[c]{@{}l@{}}Some high   school: .6\%\\    High school: 11.7\%\\    Some college: 12.1\%\\    College: 47.3\%\\    Postgraduate: 28.3\%\end{tabular}
\end{tabular}
\caption{Demographic breakdown of survey participants in each month of data collection.}\label{tab:monthly}
\end{table}

\section{Survey Participants' Data Distributions}\label{sec:demo}
We excluded a total of 30 participants for providing an invalid age (e.g., below 18, over 100; $n = 9$) and for completing the survey in an unusually short time (e.g., under three minutes; $n = 21$). The weighting benchmarks for ensuring that our sample is representative were drawn from \citet{mercer2018selection}'s comprehensive framework based on nine population datasets of the United States, and updated according to Pew Research’s 2020 study of the American Electorate \cite{gramlich2020electorate}. 

The final sample included 12,903 individuals, of whom 46\% were men ($n = 5,929$), 52.3\% were women ($n = 6,748$), and 1.8\% were non-binary ($n = 226$). On average, our sample was relatively diverse: 66.2\% identified  as white ($n = 8,536$), 14.2\% as Black or African American ($n = 1,838$), 8.8\%  as Asian or Asian American ($n = 1,135$), 7.4\% as Hispanic or Latino ($n =  958$), .6\% as American Indian or Native American ($n = 79$), .2\% as Pacific Islander or Alaska Native ($n = 30$), and 2.5\% as another race or ethnicity ($n 
= 327$). They represented a broad range of ages, ranging from 18 to 98, with an average age of 37.6 years old (SD = 11 years). The sample was relatively well-educated: approximately 15.4\% attended some college ($n = 1992$), 9\% received their Associate’s degree ($n = 1177$), 44.2\% received their Bachelor’s degree ($n = 5705$), 18.3\% received their Master’s degree ($n = 2359$), 2.5\% received their Doctorate ($n = 324$) and 2.6\% completed a professional school ($n = 340$),  7.5\% completed high school ($n = 964$) and .3\% completed some high school ($n = 42$). 
The month-by-month demographics are in Table \ref{tab:monthly}.
\section{Survey Item Details}\label{sec:surv}
To measure \textit{trust}, the survey items assessed people's perceptions of 1) \textit{AI's ability} to achieve core tasks effectively (``I believe that AI will produce output that is accurate"), 2) \textit{AI's benevolence} to produce positive outcomes for the public (``I believe that AI will have a positive effect on most people"), and 3) \textit{AI's integrity} in facilitating user safety (``I believe that AI will be safe and secure to use."). All items were scaled on a 5-point Likert scale (1 = strongly disagree, 5 = strongly agree) and composited into a trust index in line with \cite{jakesch2019ai, ma2017self}.

For \textit{willingness to adopt AI}, participants were asked to indicate how willing they would be to rely on information provided by AI, depend on decisions made by AI, share relevant information about themselves to enable AI systems to do tasks for them, allow their data to be used by AI, and share their feelings with AI.  Items were scored on a 7-point Likert scale (1 = completely unwilling, 7 = completely willing), and composited into averages in line with past work by \cite{kelly2023factors}.  
\section{Removing potentially AI-generated responses}\label{sec:exclu}
 To filter out potentially AI-generated responses, we first queried ChatGPT 100 times using variations of the prompt we gave to the participants, thus obtaining 100 responses $m^{GPT}_i$, where $i = 1, .., 100$. For each GPT-generated response, we used the sentence embedding model \texttt{all-mpnet-base-v2} \cite{reimers-gurevych-2019-sentence} to generate a 768-dimensional embedding  $e(m^{GPT}_i)$, thus constructing a set of embeddings $E^{GPT} = \{e(m^{GPT}_i) | i = 1, ..., 100\}$ representing AI-generated metaphors.  We also computed a embedding $e({m_p})$ for each participant-written metaphor $m_p$. Then, for each participant-written metaphor, we measure the cosine similarity $sim(e({m_p}), e(m^{GPT}_i)$ between $e({m_p})$ and each embedding $e(m^{GPT}_i) \in E^{GPT}.$ If any cosine similarity was greater than 0.85,\footnote{We choose this threshold based on manual inspection of how \textit{semantically} similar the metaphors are at different thresholds.} we identified it as highly similar to a GPT-generated response and thus excluded it from the dataset. 

 \begin{figure*}
     \includegraphics[width=0.9\linewidth]{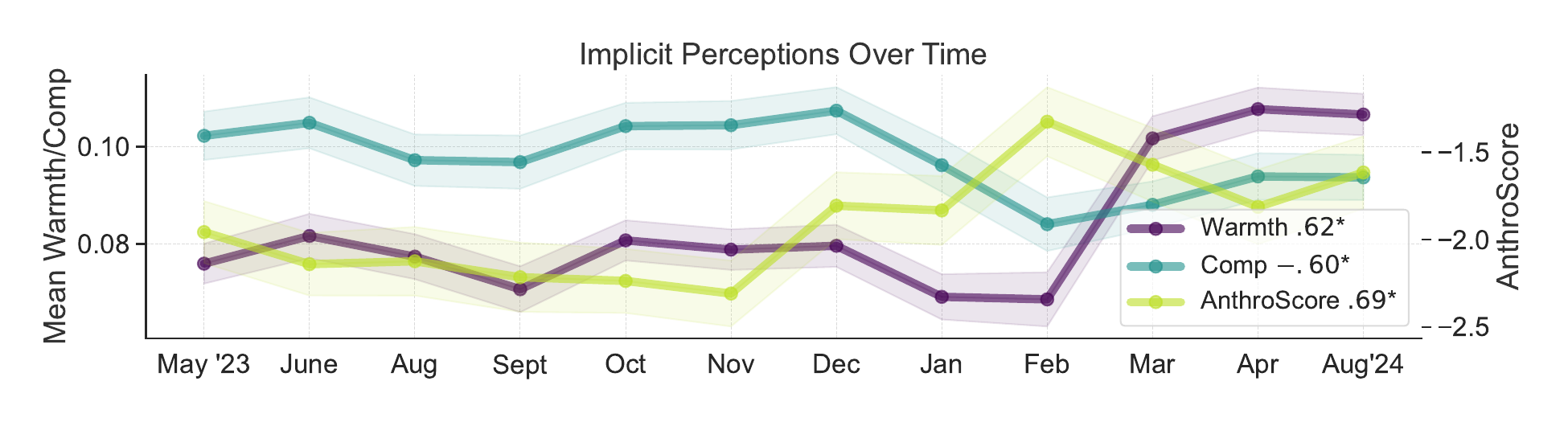}
     \caption{Version of Fig. \ref{fig:topics_over_time} (top) using AnthroScore rather than the binarized percentage form similarly reveals that anthropomorphism is increasing ($r = 0.69^*$)}
     \label{}
 \end{figure*}
\section{Further Demographic Differences in Metaphors and Perceptions}\label{sec:moredetail}
\begin{figure*}
    \centering
    \includegraphics[width=0.45\linewidth]{ 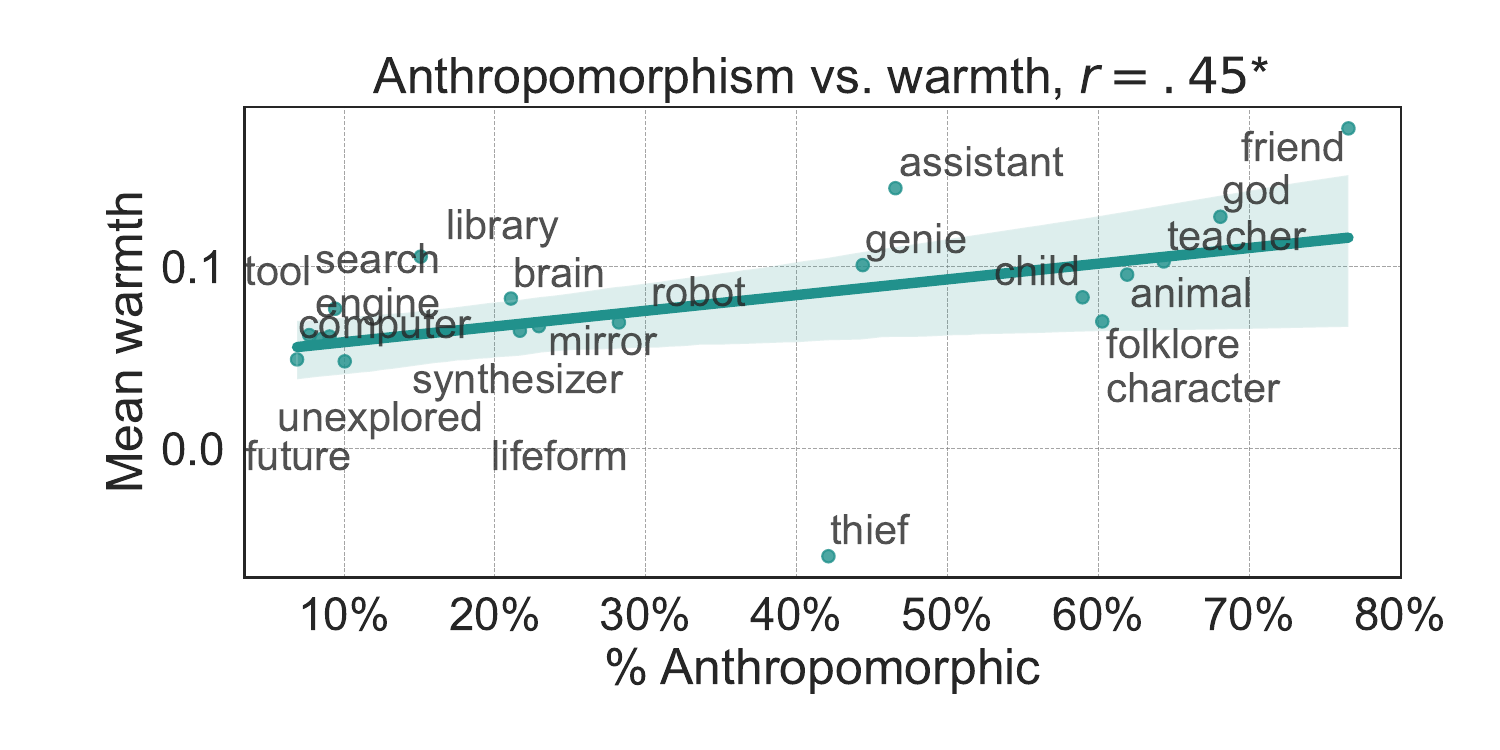}
    \includegraphics[width=0.45\linewidth]{ 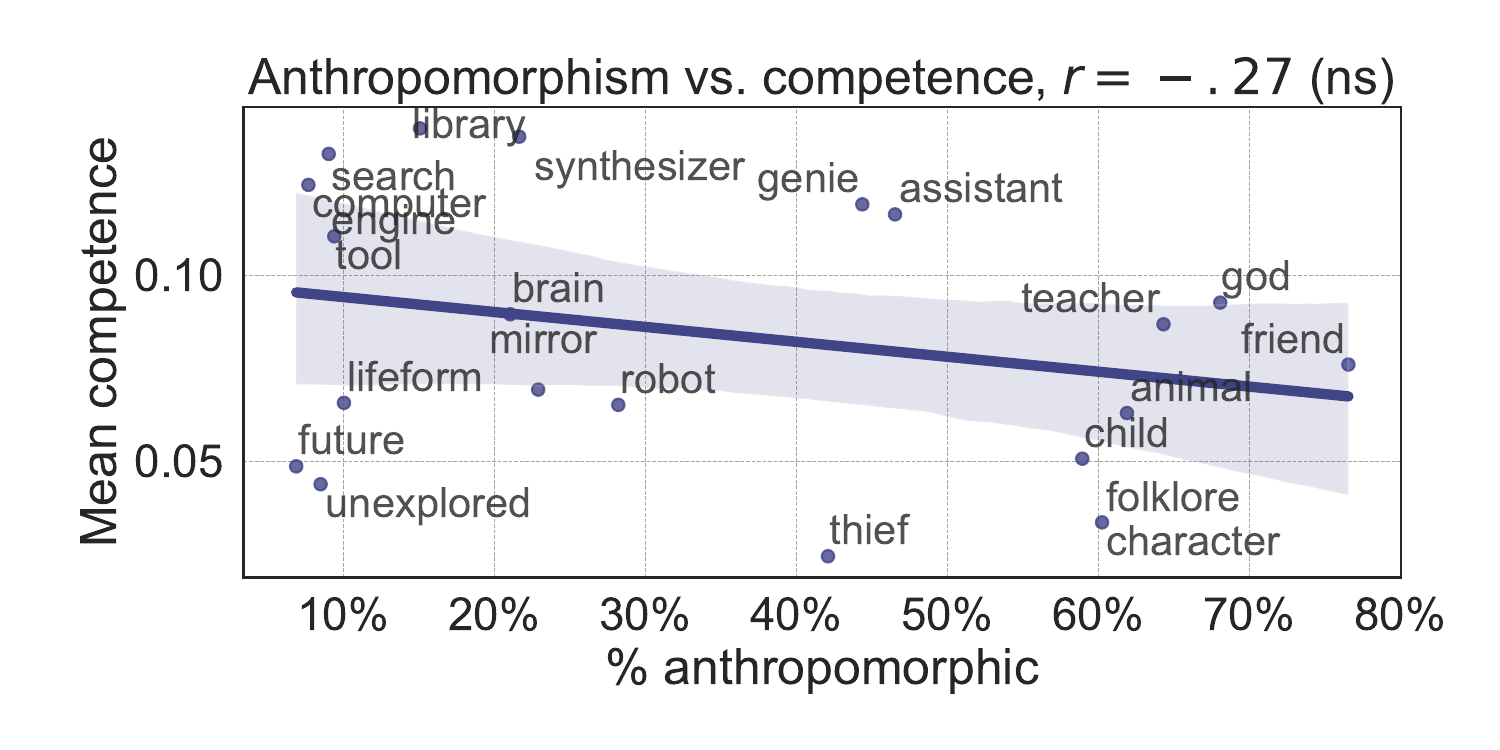}
    
    \caption{\textbf{Correlations between implicit perceptions of AI as anthropomorphic versus warm (left), versus competent (center), and temporal shifts in these implicit perceptions over time (right).} We find that anthropomorphism and warmth are significantly positively correlated ($r = 0.46, p < 0.05$), but there is no significant correlation between anthropomorphism and competence. We find that anthropomorphism and warmth are increasing over time, while competence is decreasing. This corroborates that the increase in anthropomorphism of AI is associated primarily with an increase in warmth.
}\label{fig:correlations_2}
\end{figure*}

\begin{figure*}
    \centering
    \includegraphics[width=0.45\linewidth]{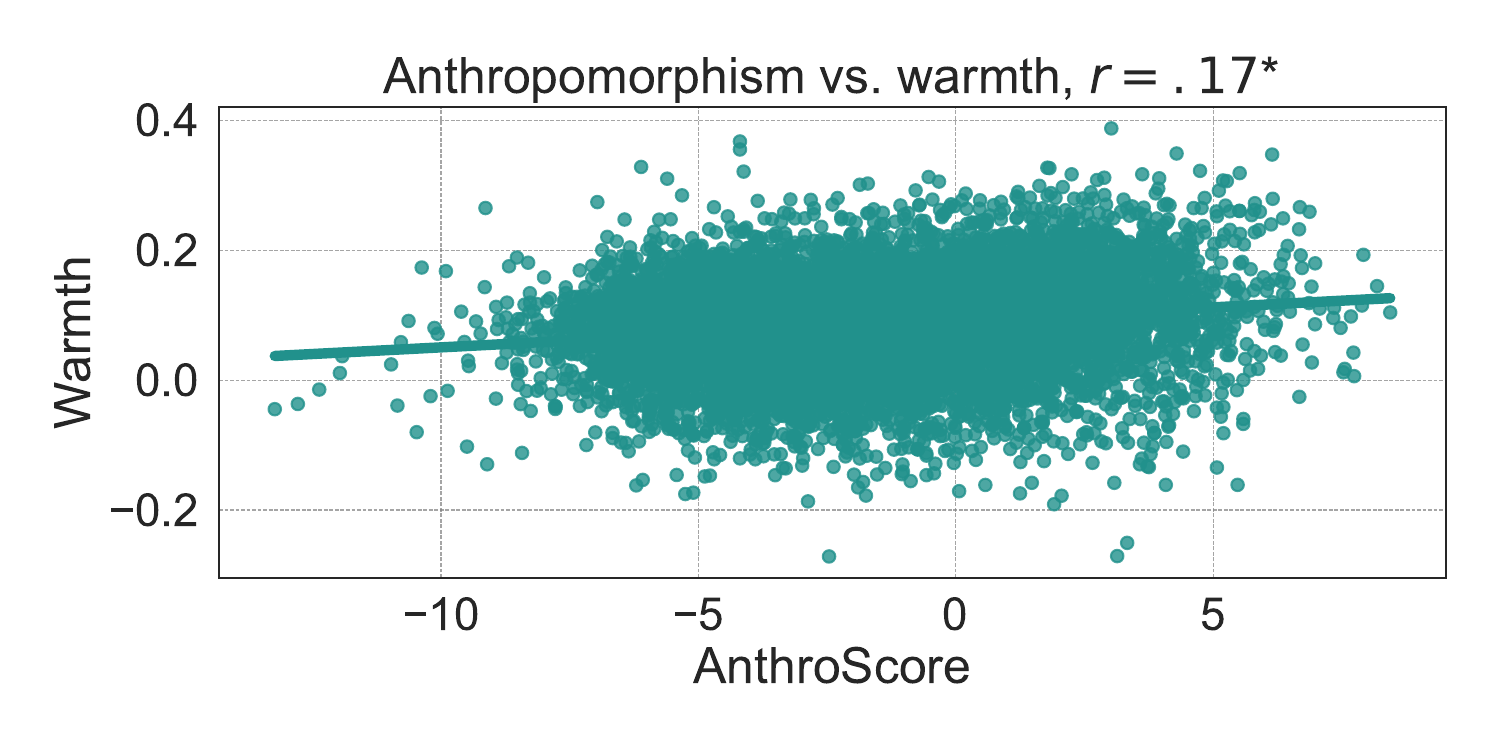}
    \includegraphics[width=0.45\linewidth]{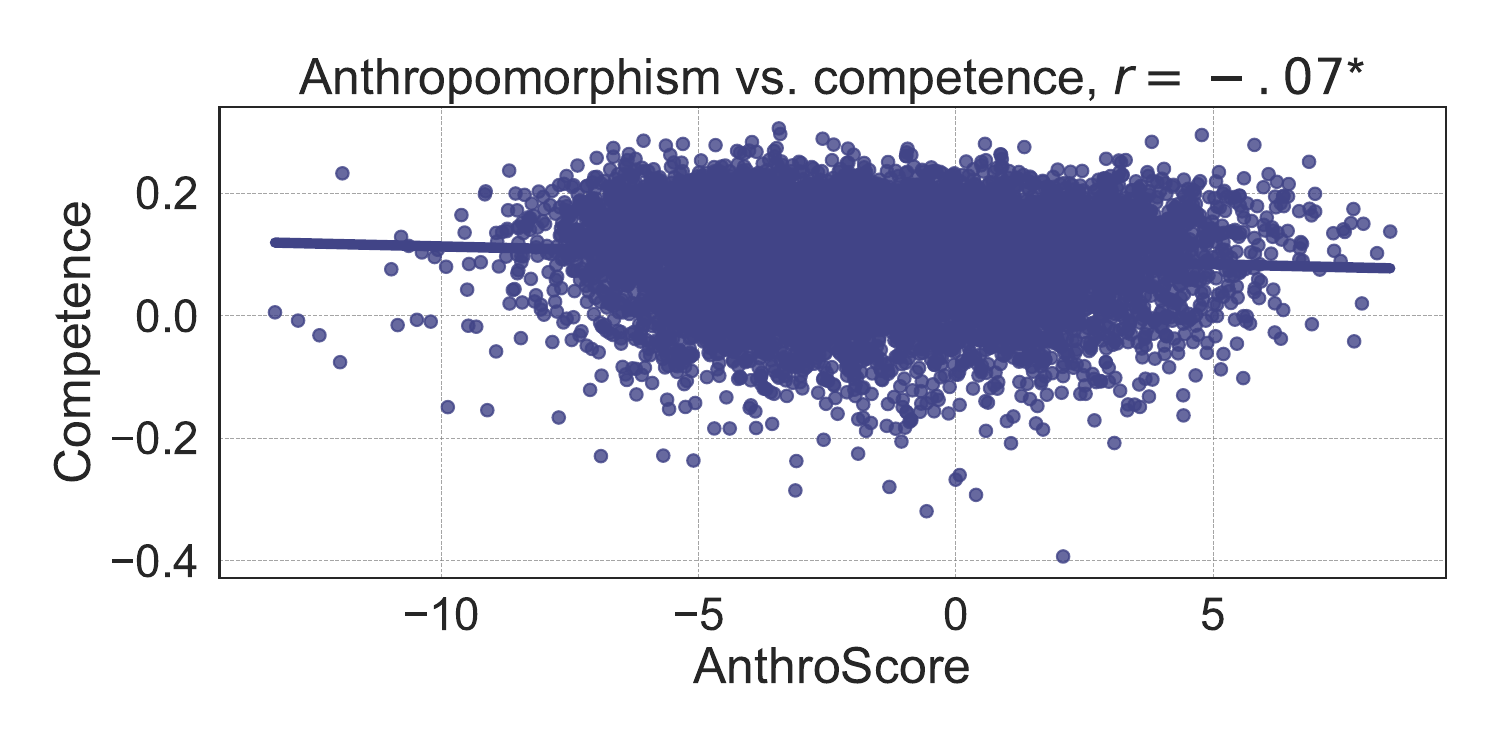}
    
    \caption{\textbf{Correlations between implicit perceptions of AI as anthropomorphic versus warm (left), versus competent (center) based on individual metaphors}. We find that anthropomorphism and warmth are weakly positively correlated ($r = 0.17, p < 0.05$), while anthropomorphism and competence are extremely weakly negatively correlated ($r = -0.07, p < 0.05$).
}\label{fig:correlations_3}
\end{figure*}
\begin{figure*}
    \centering
\includegraphics[width=0.3\linewidth]{ 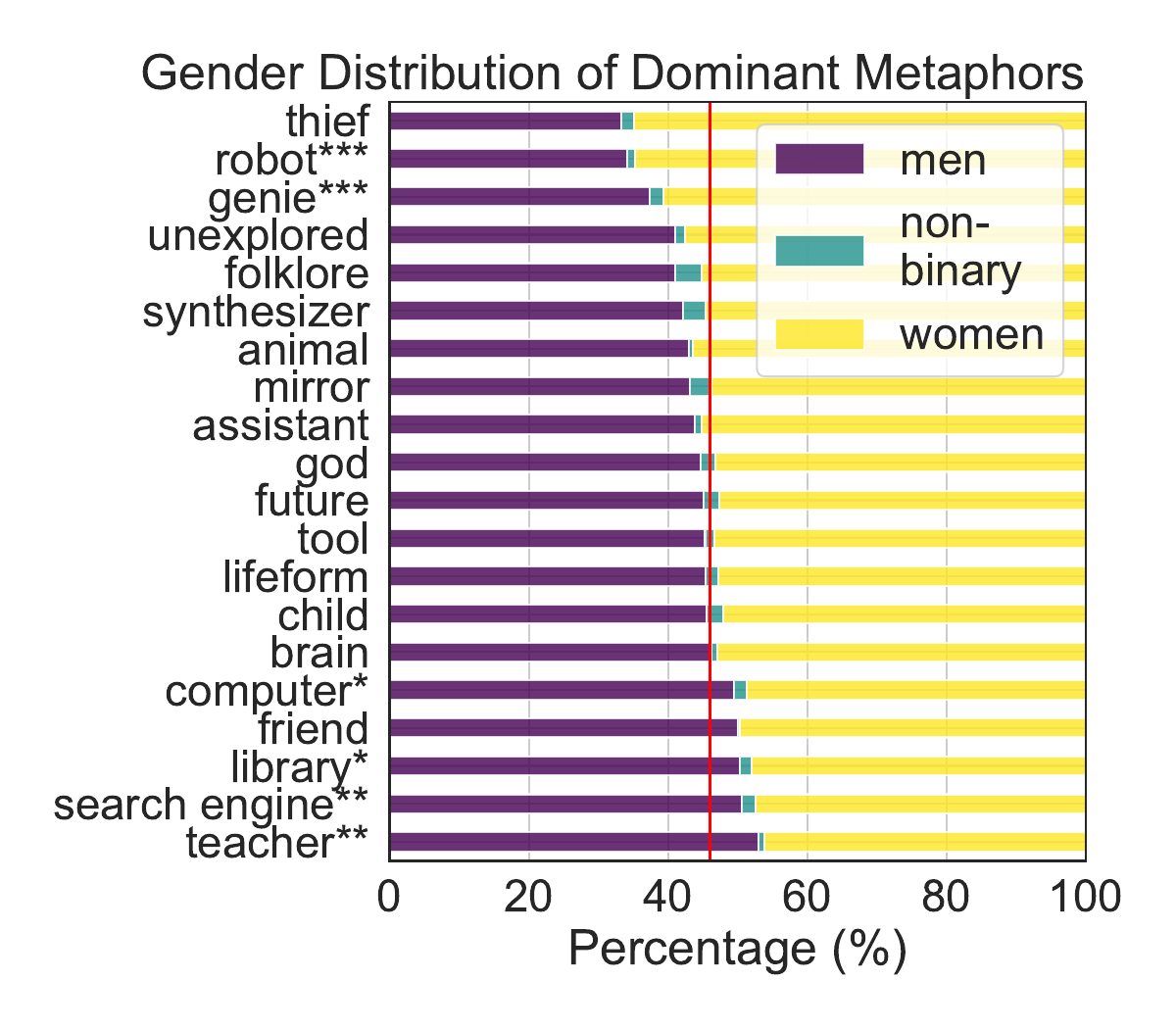}
\includegraphics[width=0.3\linewidth]{ 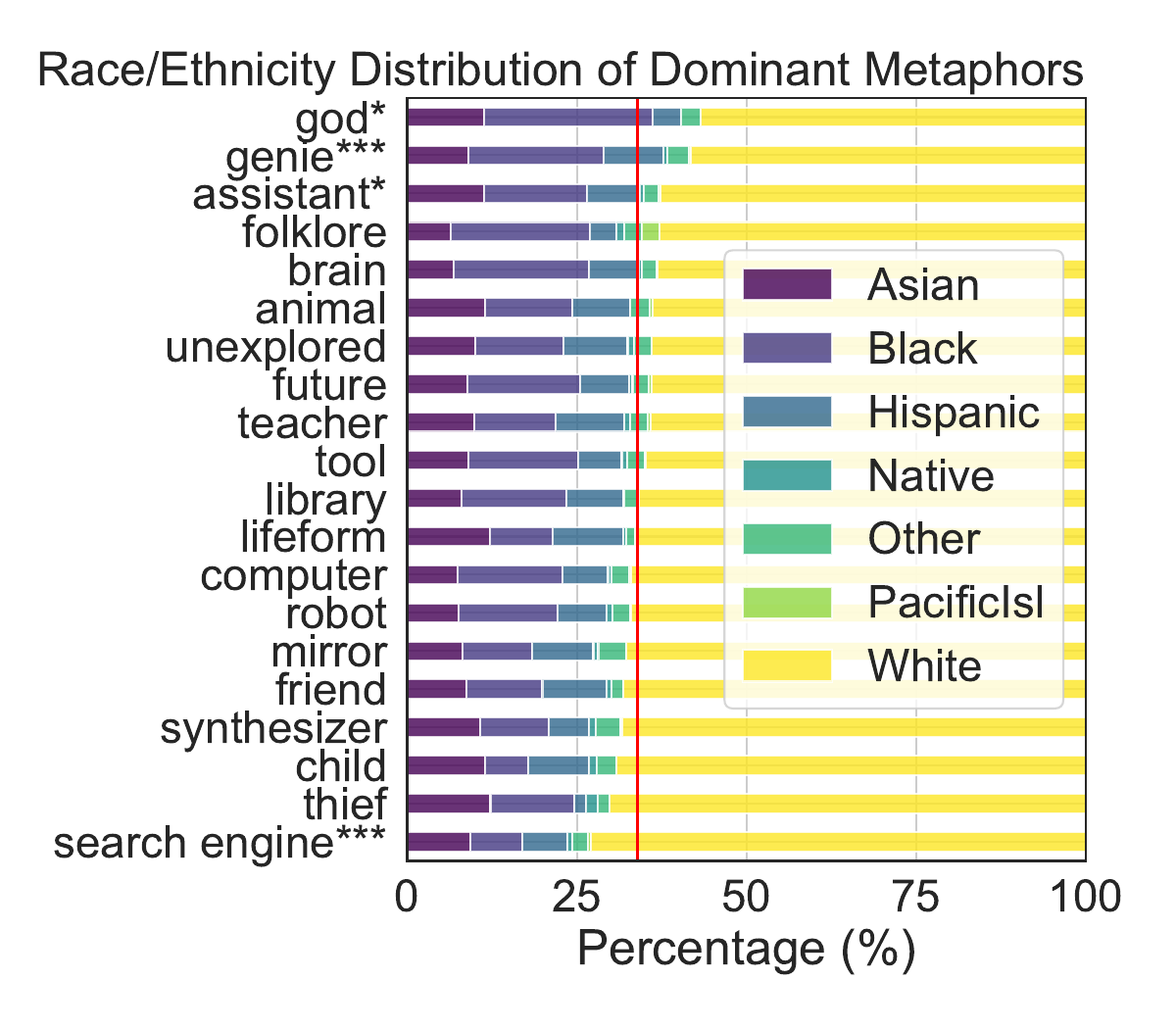}
\includegraphics[width=0.3\linewidth]{ 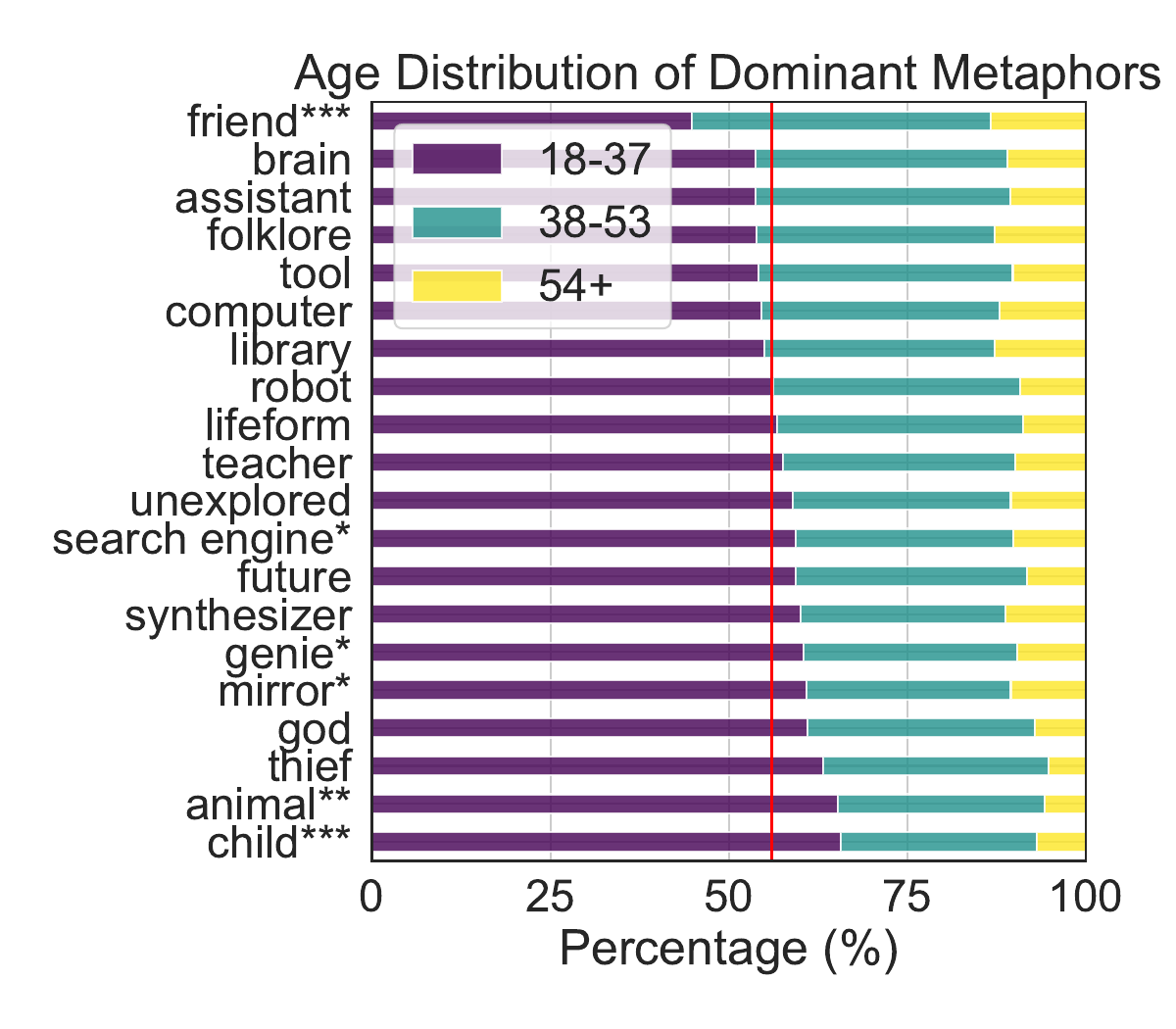}
    \caption{\textbf{Demographic differences in dominant metaphors.} Dominant metaphors labeled with *s have statistically significant demographic differences based on a binomial test (between men and non-men for gender; between white and non-white for race/ethnicity; between 18-37 and 37+ for age), and the red line indicates the expected demographic proportion under the null hypothesis.}
    \label{fig:dem_diff_metaphors}
\end{figure*}
\begin{figure*}
    \centering
\includegraphics[width=0.8\linewidth]{ 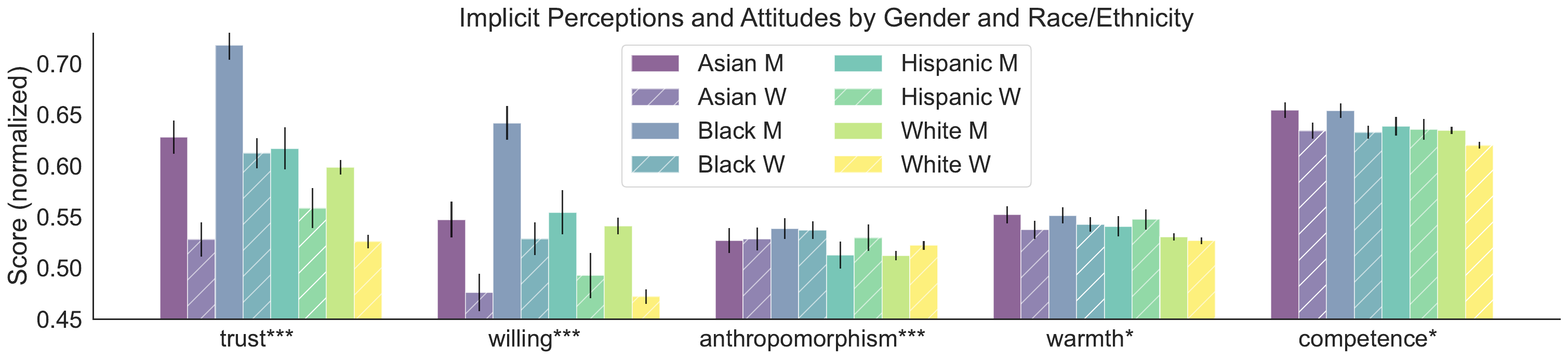}
    \caption{\textbf{Intersectional analyses of implicit perceptions.}}
    \label{fig:intersec}
\end{figure*}
\begin{figure*}
    \centering
\includegraphics[width=0.8\linewidth]{ 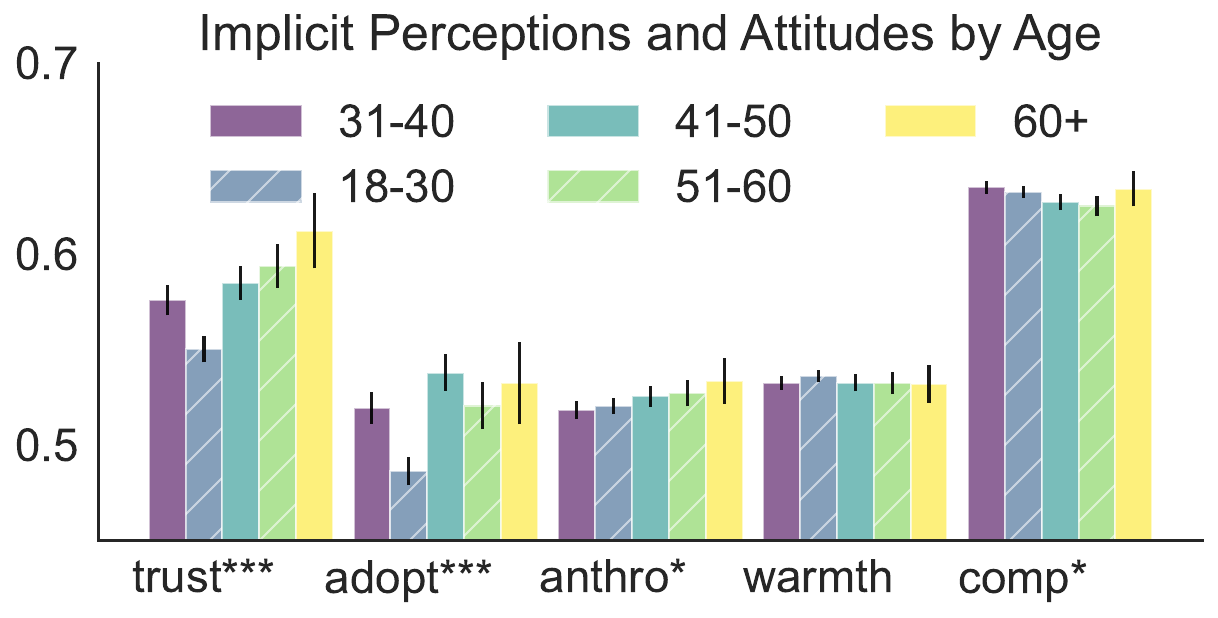}
\includegraphics[width=0.8\linewidth]{ 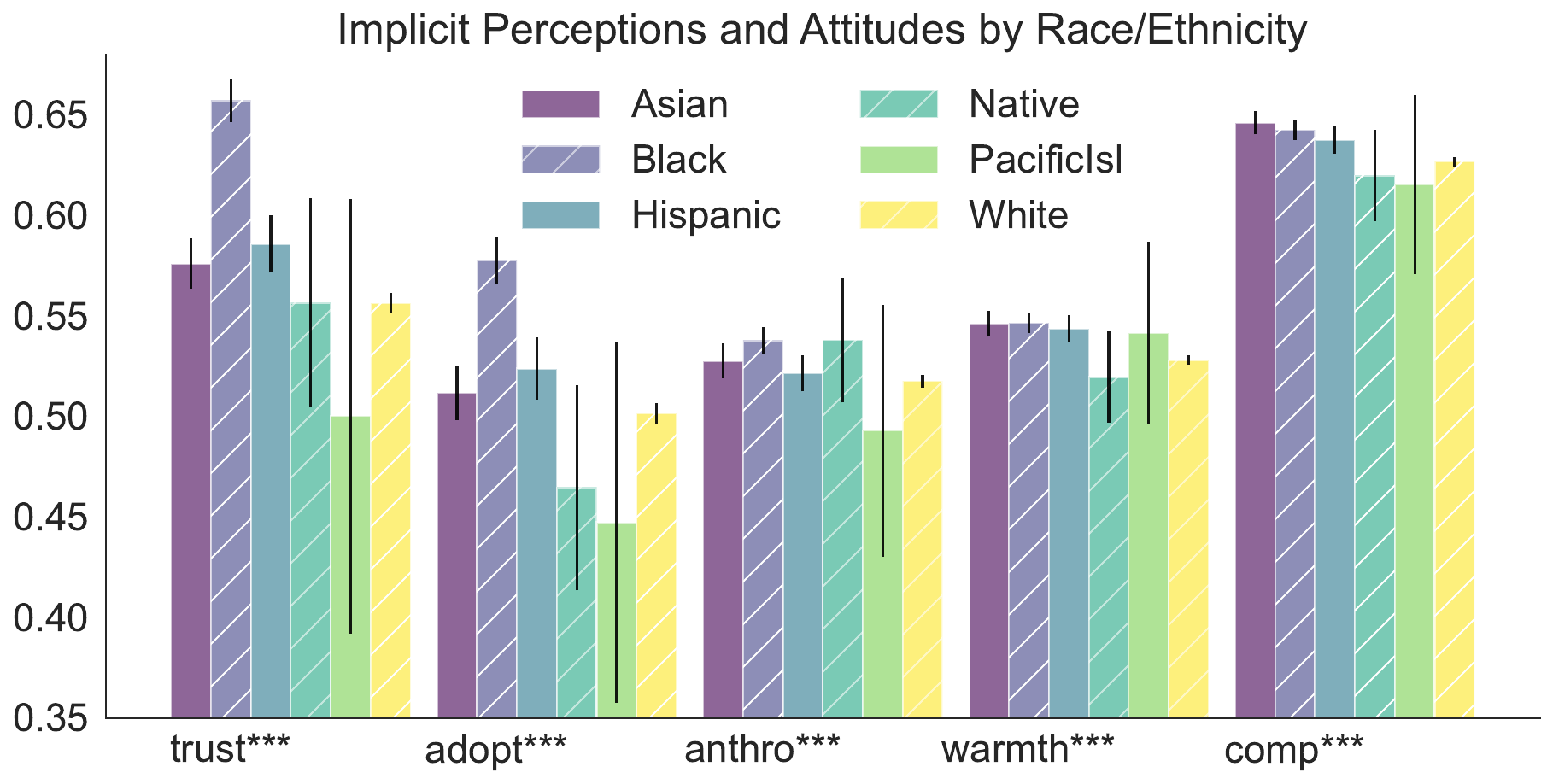}
    \caption{\textbf{Implicit perceptions and atititudes by age and race/ethnicity, extended from the main text -- including 10-year age brackets and full distributions of race/ethnicities respectively. These plots reflect similar patterns as described in the main text.}}
    \label{fig:race2}
\end{figure*}

Implicit perceptions and attitudes by race/ethnicity and 10-year age brackets are in Figure \ref{fig:race2}. Distributions of dominant metaphors by different demographics are in Figure \ref{fig:dem_diff_metaphors}.
\paragraph{Intersectional Analysis}
Table \ref{tab:intersec} highlights the top metaphors for each gender-by-race/ethnicity group. 
Figure \ref{fig:intersec} shows differences in overall rates of perceptual dimensions by gender-by-race/ethnicity group. 
\begin{table*}[]
    \centering
    \begin{tabular}{ll}
    \textbf{Race/ethnicity and gender} & \textbf{Top 3 metaphors}\\\hline
       white M & search engine, genie, robot\\\hline white W & robot, future shaper, teacher\\\hline Asian M & robot, genie, synthesizer\\\hline 
       Asian W & brain, library, assistant\\\hline 
       Hispanic W & unexplored realm, thief, tool\\\hline Hispanic M & teacher, god, unexplored realm\\\hline Black W & search engine, child, brain\\\hline Black M & brain, child, search engine\\\hline 
    \end{tabular}
    \caption{Top three metaphors for intersectional gender-by-race/ethnicity groups.}
    \label{tab:intersec}
\end{table*}
\begin{figure*}
    \centering
    \includegraphics[width=0.45\linewidth]{ 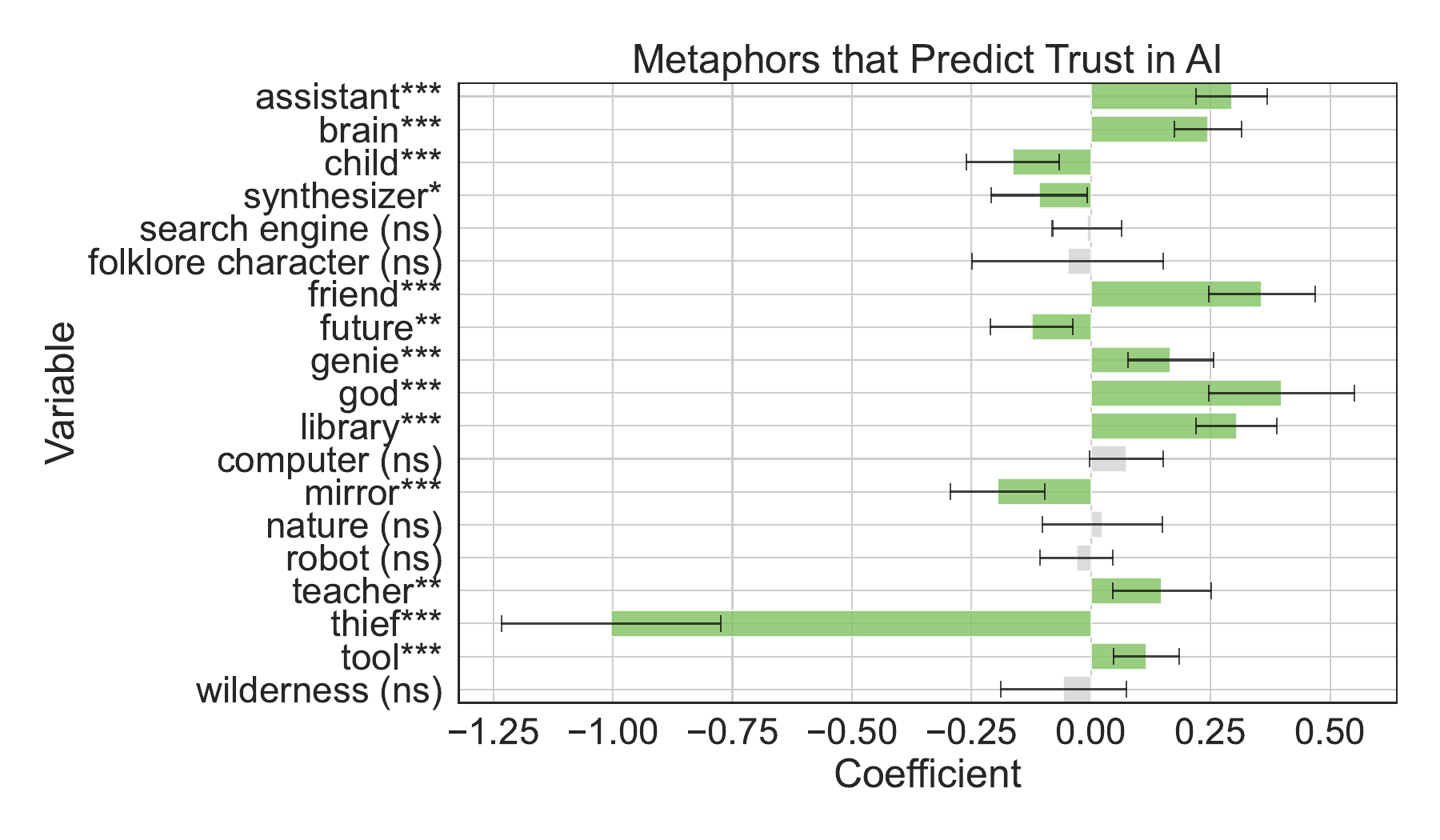}
    \includegraphics[width=0.45\linewidth]{ 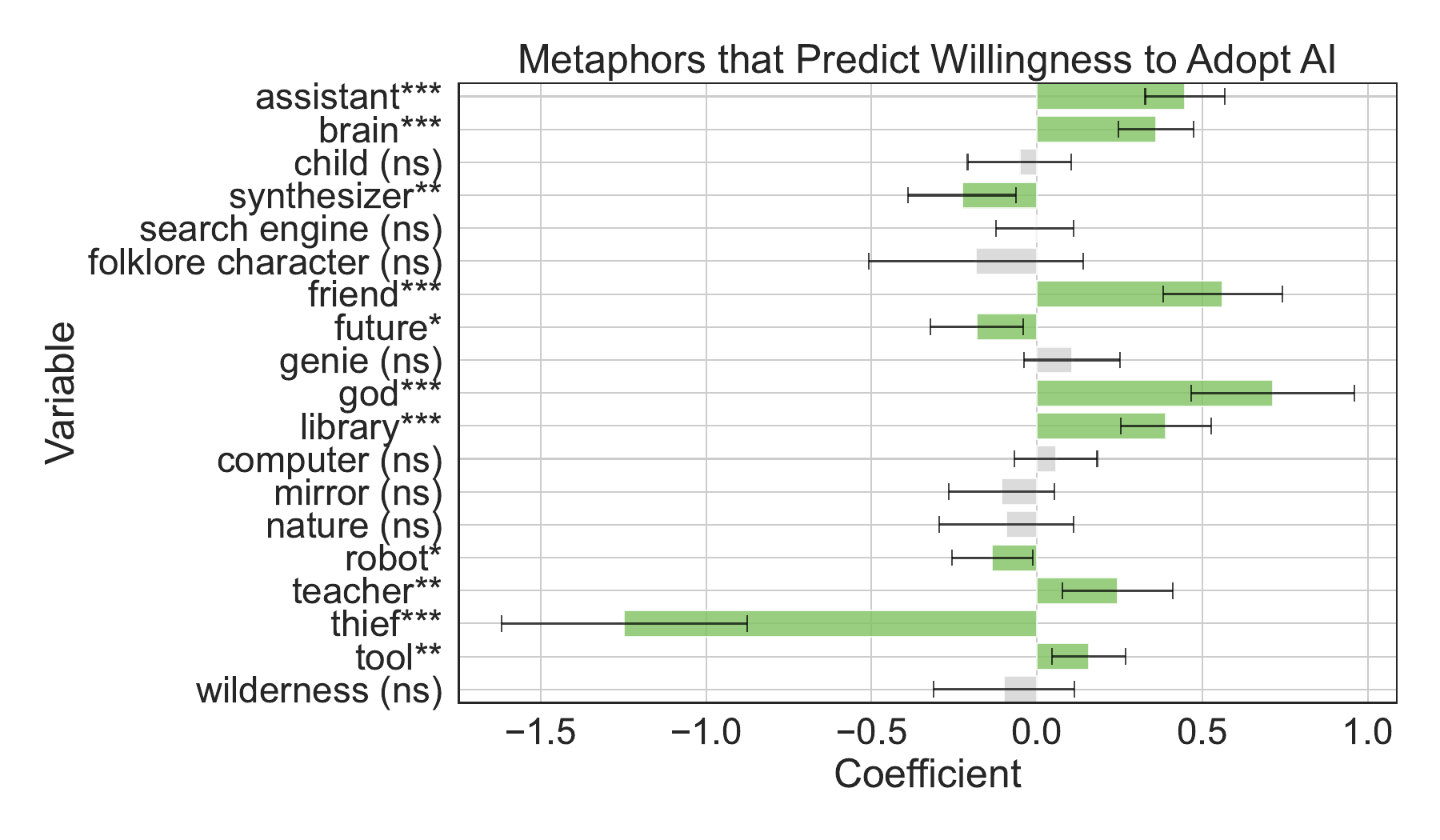}
    \includegraphics[width=0.45\linewidth]{ 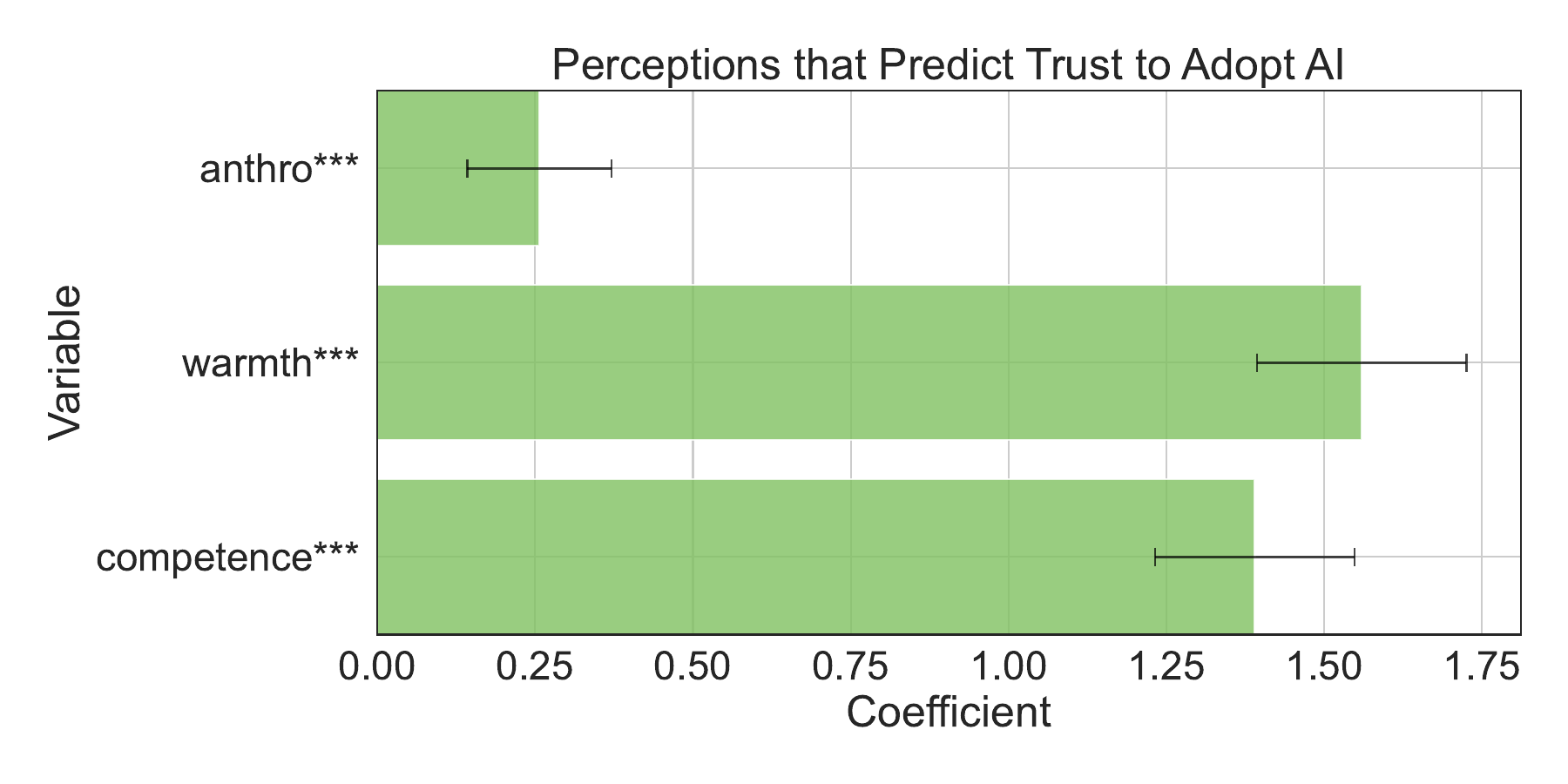}
    \includegraphics[width=0.45\linewidth]{ 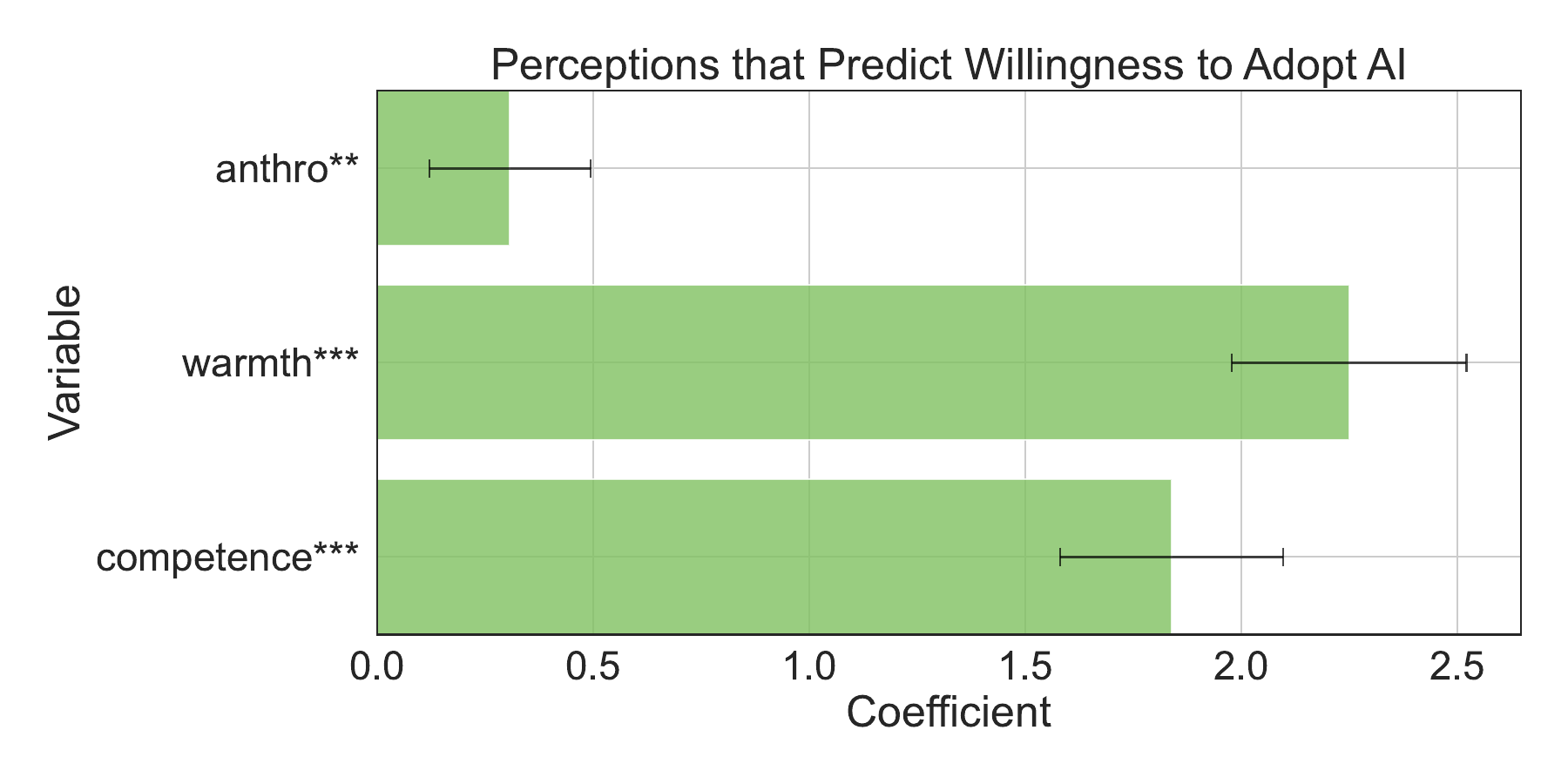}
    
    \caption{\textbf{Regressions to predict trust and adoption from dominant metaphors alone (top) and from implicit perceptions alone (bottom).} Error bars represent 95\% CI.}
    \label{fig:regressions}
\end{figure*}
\tiny 
\begin{table*}
\begin{tabular}{|p{0.05\linewidth}p{0.93\linewidth}|}
\hline
\textbf{Metaphor}        & \textbf{Examples}                                                                                                                                 \\\hline animal & ``AI is trained by feeding in a large amount of text (or images, sound, depending upon the type of AI).  It learns patterns in the text and then is able to parrot back information, based on what it has seen.'' ``AI is like a dog. You train it and it does what it thinks you want it to do.'' ``AI is like a dog, you can train it to do one thing very specifically well and then when it 100 times out of 100 gives you the expected output for the task you can ask it to do something different, and again tell it what is good and what is bad. after that you can start asking it more random stuff and trust more because the dog has learned some rules that it self governs by.'' ``AI is like brilliant companion pet because it wants to please you as best it can with how it is able to understand the world.'' ``Humans and AI are just different breeds of parrots.\\\hline 
assistant & ``A metaphor would be a maid. I think AI is best used as an assistant however they should not be "the boss" they need someone to oversee them and ensure accuracy to the unique situations'' ``AI is like an assistant because it can complete tasks that you direct it to'' ``AI is like Having a personal assistant because Because it is helpful and fast response'' ``AI works like a person that is very fast and efficient and has been trained through trial and error to complete their work tasks upon some specific criteria. For example, a line worker in a factory that has been trained to recognize parts coming down the line and binning them based on their quality.'' ``I have always wanted something like alexa for the house.  More of a centralized version you see in sci fi movies.  I think of AI as being similar to that.\\\hline 
brain & ``AI is like like a computer with a functioning brain because It takes information and can recognize patterns and create new advice.'' ``AI is like an electronic brain.'' ``AI is like a floating brain because it has knowledge and is just there'' ``Mediocre result of human brain cloning'' ``That would be work brain and that is because it help to make work more easily and increases productivity.\\\hline 
child & ``AI is like a todler because it is still very young however it will grow older'' ``AI for humans is like a mother raising her child and her child grows up to be the president.'' ``AI is like A kid with a really good memory because it just soaks in all information and just spits it back at you whether it's wrong or right.'' ``AI is like an intelligent baby because it knows everything, but also needs to continuously learn'' ``AI is like A fast developing child because it is new to the things it learns but is gaining knowledge and learning at an incredible speed.''\\\hline 
computer & ``AI is like a machine I input thoughts into and gain ideas.'' ``I know that AI is a complex computer program that is supposed to “learn”, but I don’t think anything can replace an organic human interaction.'' ``AI technology in my opinion is almost like a substitute teacher. Someone who has knowledge but less personal experience with its students. So there could be a disconnect.'' ``A computer is programmed to have responses that mimic human behavior.  The AI can solve complicated problems and questions.'' ``AI is computers making answers for you.''\\\hline 
folklore & ``AI is a wolf in sheep's clothing. It looks innocent, but I think it will have dire consequenses.'' ``AI is like a fairy god-mother. She grants all your wishes when you just ask.'' ``The Jewish concept of a "golem" is extremely accurate.  It imitates life, but has none of its own, it's animated by the will and desires of its creator and not its own intelligence or will, and it will inevitably turn on its creator and destroy them.'' ``The AI is like a fairy Godmother. I say this because like a fairy Godmother allows you to obtain and push you to your goals in order to succeed the AI does that as well. The AI will give you advice and tips in order to succeed in what you're doing and since it has numerous answers to whatever question you may want resolved it can guide you towards that. It can help you find ways to study better, how to improve your intelligence, how to better your mental and physical health, how to start a new hobby, how to start a new side gig so many options to help you succeed in different aspects of your life.'' ``AI is like The anti-christ because Oh boy! It’s our friend! We love it but eventually it will surpass us to such a degree it will become our overlord, then not need us at all, making decisions it deems best but are not best for us''\\\hline 
friend & ``AI is like A know it all friend because It only gives its opinion without suggestion'' ``AI is like The one you have a love-hate relationship wirh because They are “there” for you and help you when you need but don’t really care about  you'' ``Ai is like that one friend who knows everything no matter how obscure it is and gives you a response'' ``AI is like an amazingly knowledgeable (yet sometimes inaccurate) friend whom you can consult on just about any topic because it can answer countless questions in detail, but occasionally makes up some of those details.'' ``AI is like A smart friend because they know the answers to the things that I don't''\\\hline 
future shaper & ``AI is the industrial revolution'' ``AI makes me think of those push button homes that people in the 50s envisioned the future to be like. It’s so much automation and access to information, but we don’t just have to use a mechanical interface to access the information.'' ``AI is a runaway train bringing materials across the world, we can't stop it at this point but don't know if it's going too fast and might crash into our towns and blow them up. The original plan was for good but we don't know where its going.'' ``AI is the new light bulb (as in Edison's first light bulb). It is going to transform our society and our experience of living as we have known it.'' ``It's like the internet itself. It can be used for good, but it can also be used for evil.''\\\hline 
genie & ``AI is like the man behind the curtain in The Wizard of Oz because it's going to learn and evolve itself into being something that eventually controls everything but it will all be from "behind the scenes" and everyone will blame each other for when things start going wrong and lines become more and more blurred as a result.'' ``AI can be likened to a wizard in a vast library of spells.'' ``Genie'' ``AI is like the genie in the bottle in Aladdin. It does what it is told no matter the consequences'' ``AI is like Magic because You just ask it to do something and it does it quickly and usually accurately.''\\\hline 
god & ``AI is like the biggest threat to our existence besides nuclear proliferation because we are severely underestimating how powerful Godlike intelligence really is.'' ``As if it were omnipresent.'' ``An AI to me is like that invisible all knowing object, that sees and knows everything.'' ``AI is like an all-knowing god. It can pull information from the collective human consciousness.'' ``AI is like Demigod because it knows everything''\\\hline 
library & ``AI is like A very good, indexable encyclopedia because you can look up almost anything'' ``A metaphor I would use for how AI works is like an adaptive Encyclopedia that takes its information from every corner of the internet.'' ``AI is like A vast, intricate library where every book is constantly being written and revised in real-time. because just as a library houses a wealth of information and insights that can be accessed and utilized in various ways, AI taps into a broad spectrum of data to generate responses and solutions. Like a librarian who helps you find the right book or piece of information, AI processes and organizes information to assist with diverse queries. And just as a library's collection grows and evolves, AI continuously learns and updates its knowledge base to improve its responses and adapt to new challenges.'' ``I'd like having an encyclopedia friend.'' ``AI is like a magical library because it contains most knowledge\\\hline 
lifeform & 
``AI is like a mechanical flower that grows and adapts to its environment'' ``I think AI is like an apple. We expect apples to be shiny and somewhat round and sweet. However, not all apples are perfectly round, and some may have bug holes too. In other words, AI can be expected to be reliable, but it is wrong sometimes too.'' ``AI works like a Bonsai tree. Humans train and grow the Bonsai tree in a particular way that replicates a full-grown tree in nature. The Bonsai tree is restricted to grow inside of the miniature pot that it is placed in, and cannot grow outside of it. It relies on a human to prune and tend to it, much like AI needs a human to program, develop and eventually utilize it.'' ``It reminds me of a living web--like something being constantly woven by a spider who moves impossibly fast. AI currently function most by forging new connections.'' ``AI is like a winding river, flowing with information sourced from deep underground. Sometimes the water rages and splashes like erroneous answers, other times it cuts and carves landscapes like automation of systems, and yet other times it flows still and calm like guided accurate responses.'' 
\\\hline 
mirror & ``AI is like the magic mirror from Snow White, it can give you answers but can at times be vague'' 
``AI is an amateur writer. It can look at others' work and make an approximate recreation, but has ability to accurately evaluate its outputs because it cannot critically think about why it wrote what it did.'' ``AI is a mimic that has learned what words are supposed to go together for a given topic prompt. It is like a cargo cult. They can make the symbols of the planes, the crates, the runway, and the storage buildings, but they have no idea how any of it works or what it means.'' ``AI is like a mirror that is able to reflect the content of humanity that is has been exposed to in order to reflect back an answer to an inquiry or task.'' ``A mirror. AI just reflects back the worst of our humanity and the people that create it. Everything that already exists on Google is just being used to feed AI.''\\\hline 
robot & ``I think it would be similar to a robot doing manual labor. Like the robots that fill my orders at amazon.'' ``AI is like a robot because it's all computer generated, artificial intelligence.'' ``Smart robots'' ``Using AI is like using a personal robotic assistant, or like a smart fake human.'' ``At this point, to me, AI is the little Wall-e character who is trying very hard to do the right thing; however, as we all know, life isn't that cut and dried. There are nuances to problems that I think AI will never be able to answer.''\\\hline 
search engine & ``It takes data and uses it to predict and modifies those predictions based on interactions and how well it performed in the past.  machine learning.'' ``It is like using all of the internet all at once. Grasping all the good parts on the knowledge you can find into a single machine that can help you in an instant'' ``AI is like a smart search engine because you plug in queries and come up with randomized answers'' ``It's like a virtual detective doing all the work for us. It sifts through all the info online and gives us the best answers.'' ``AI is like a search engine because that's all it actually is''\\\hline 
synthesizer & ``AI is like a piece of clay that is molded by multiple hands. because of the fact that AI models get their information from multiple different sources, whether it be stable diffusion with different artists or ChatGPT searching through the internet for information.'' ``It's kinda like those He-Man action figures. It was all the same five molds, but in different combinations and colors. Most of them were very ugly, but they all were different. I think input really matters with AI.'' ``AI is like a coded muse because it fleshes out ideas, plans, and art almost instantaneously'' ``Its like putting together a quilt. You make each square individually with different thoughts then you string, or sew them all together to create a fully formed quilt/picture.'' ``It is like making a movie. You give a different prompt, these are like the actors and scripts. Then depending on the script and actors you will get a certain movie.''\\\hline 
teacher & ``AI is like a trusted professor because it always has the right answer'' ``AI is like a smart person that can do anything for you without emotions involved because it is just technology but enhanced'' ``I cannot think of anything appropriate, other than that AI is akin to a learned person in whichever field the question fits into. When it comes to art and more complex requests, I have my doubts'' ``AI is like that family member who believes they know everything, even though it's not the case'' ``AI is like the really smart stoner kid that comes up with some amazing ideas innovative ideas, but are not always aligned with the conversation.''\\\hline 
thief & ``AI is a grave robber, stealing from others and calling the output it's own.'' ``taking over jobs'' ``Job stealing'' ``AI is what China does to big name industry fashion brands, but when I say that, I mean how reps were back in the day 20 years ago like jordans but the logo has his meat out'' ``Taking people’s jobs''\\\hline 
tool & ``A calculator'' ``AI helps you do certain manual labor work such as creating a spread sheet of data in seconds. By using AI a person can save time using it while being able to do other tasks'' ``AI is like tools for work because it can be an aid but also but a hinderance if it isn't up to date.'' ``AI is like A tool because It depends on how you use it'' ``A hammer that can either be used to build or destroy things.''\\\hline 
unexplored realm & AI is somewhat like a tsunami, it hit suddenly and spread out fast.'' ``I will liken AI to the entire milky way with so much wonder and beauty. With the planets and celestial bodies that make up the Milky Way being the varied components put together to power the AI as we know it'' ``like a black hole'' ``AI is an open door to the world we live in.'' ``AI is like outerspace because there's so much we don't know about it\\\hline
\end{tabular}
\caption{\textbf{Additional examples of participants' responses in each dominant metaphor cluster.}} \label{tab:metaphor_examples}
\end{table*}

\begin{table*}[]\small
\begin{tabular}{|p{0.1\linewidth}llllll|}\hline
                                        & \multicolumn{3}{c}{\textbf{Trust in AI}}                                            & \multicolumn{3}{|c|}{\textbf{Willingness to adopt AI}}      \\\hline
                                        & $\beta$          & adjusted $r^2$ & $\Delta r^2$ & $\beta$         & adjusted $r^2$ & $\Delta r^2$ \\\hline
\textbf{Block 1 - Demographics and use} &               & 0.12                          &                            &               & 0.10                          &                            \\
frequency of AI tool use                & $1.54^{***}$  &                               &                            & $2.71^{***}$  &                               &                            \\
non-man                                 & $-0.22^{***}$ &                               &                            & $-0.28^{***}$ &                               &                            \\
age                                     & $0.38^{***}$  &                               &                            & $0.48^{***}$  &                               &                            \\
non-white                               & $0.13^{***}$  &                               &                            & $0.10^{***}$  &                               &                            \\
time                                    & $0.16^{***}$  &                               &                            & $0.35^{***}$  &                               &                            \\\hline
\textbf{Block 2 - Dominant metaphors}   &               & 0.15                          & $0.03^{***}$                       &               & 0.13                          & $0.02^{***}$                       \\
assistant                               & $0.00$ (ns)   &                               &                            & $0.04$ (ns)   &                               &                            \\
brain                                   & $0.15^{***}$  &                               &                            & $0.23^{***}$  &                               &                            \\
child                                   & $-0.17^{***}$ &                               &                            & $-0.06$ (ns)  &                               &                            \\
synthesizer                             & $-0.22^{***}$ &                               &                            & $-0.39^{***}$ &                               &                            \\
search engine                           & $-0.10^{**}$  &                               &                            & $-0.15^{*}$   &                               &                            \\
folklore character                      & $-0.05$ (ns)  &                               &                            & $-0.18$ (ns)  &                               &                            \\
friend                                  & $0.03$ (ns)   &                               &                            & $0.11$ (ns)   &                               &                            \\
future                                  & $-0.03$ (ns)  &                               &                            & $-0.06$ (ns)  &                               &                            \\
genie                                   & $-0.02$ (ns)  &                               &                            & $-0.15^{*}$   &                               &                            \\
god                                     & $0.16^{*}$    &                               &                            & $0.40^{***}$  &                               &                            \\
library                                 & $0.04$ (ns)   &                               &                            & $0.00$ (ns)   &                               &                            \\
computer                                & $-0.00$ (ns)  &                               &                            & $-0.05$ (ns)  &                               &                            \\
mirror                                  & $-0.17^{***}$ &                               &                            & $-0.07$ (ns)  &                               &                            \\
lifeform                                  & $0.08$ (ns)   &                               &                            & $-0.03$ (ns)  &                               &                            \\
robot                                   & $0.02$ (ns)   &                               &                            & $-0.07$ (ns)  &                               &                            \\
teacher                                 & $-0.04$ (ns)  &                               &                            & $-0.02$ (ns)  &                               &                            \\
thief                                   & $-0.60^{***}$ &                               &                            & $-0.69^{***}$ &                               &                            \\
tool                                    & $0.01$ (ns)   &                               &                            & $-0.00$ (ns)  &                               &                            \\
wilderness                              & $0.01$ (ns)   &                               &                            & $-0.03$ (ns)  &                               &                            \\\hline
\textbf{Block 3 - Implicit perceptions} &               & 0.21                          & $0.06^{***}$                      &               & 0.18                          & $0.05^{***}$                        \\
anthropomorphism                        & $0.24^{***}$  &                               &                            & $0.20^{*}$    &                               &                            \\
warmth                                  & $1.26^{***}$  &                               &                            & $1.68^{***}$  &                               &                            \\
competence                              & $1.23^{***}$  &                               &                            & $1.72^{***}$  &                               &    \\\hline                       
\end{tabular}
\caption{\textbf{Hierarchical regression results of demographic and use variables, dominant metaphors, and implicit perceptions on trust in AI and willingness to adopt AI.} Statistical significance of $\Delta r^2$ values are determined using an F-test.}\label{tab:regressioncoef}
\end{table*}
\end{document}